\documentclass[10pt]{article}
\usepackage{amssymb}
\usepackage{amsfonts}
\usepackage{amsmath}
\newcommand{\rd}{}
\newcommand{\gn}{}

\numberwithin{equation}{section}

\newcommand{\bC}{\mathbb{C}}

\newcommand{\bN}{\mathbb{N}}
\newcommand{\bQ}{\mathbb{Q}}

\newcommand{\bR}{\mathbb{R}}
\newcommand{\bT}{\mathbb{T}}

\newcommand{\kB}{{\mathcal B}}
\newcommand{\kD}{{\mathcal D}}

\newcommand{\kG}{{\mathcal G}}

\newcommand{\kI}{{\mathcal I}}
\newcommand{\kJ}{{\mathcal J}}

\newcommand{\kU}{{\mathcal U}}

\newcommand{\ga}{{\alpha}}
\newcommand{\gb}{{\beta}}
\newcommand{\gd}{{\delta}}
\newcommand{\gD}{{\Delta}}
\newcommand{\gga}{{\gamma}}
\newcommand{\gG}{{\Gamma}}
\newcommand{\gk}{{\kappa}}
\newcommand{\gL}{{\Lambda}}

\newcommand{\gr}{{\varrho}}
\newcommand{\gs}{{\sigma}}
\newcommand{\gt}{{\tau}}

\newcommand{\gth}{{\theta}}

\newcommand{\gotD}{{\mathfrak D}}

\newcommand{\gotH}{{\mathfrak H}}
\newcommand{\gotL}{{\mathfrak L}}

\newcommand{\goth}{{\mathfrak h}}
\newcommand{\gotl}{{\mathfrak l}}

\newcommand{\gotX}{{\mathfrak X}}
\newcommand{\gotY}{{\mathfrak Y}}

\newcommand{\gotZ}{{\mathfrak Z}}
\newcommand{\gotz}{{\mathfrak z}}

\newcommand{\real}{{\Re\mbox{\rm e}}}

\newcommand{\dom}{{\mbox{\rm dom}}}
\newcommand{\ran}{{\mbox{\rm ran}}}

\newcommand{\supp}{{\mbox{\rm supp}}}

\newcommand{\bv}{\vert\vert\vert}
\newcommand{\oI}{\overline{\kI}}

\newcommand{\res}{\gr}

\newtheorem{theo}{Theorem}[section]
\newtheorem{prop}[theo]{Proposition}
\newtheorem{lem}[theo]{Lemma}
\newtheorem{cor}[theo]{Corollary}

\newtheorem{defi}[theo]{Definition}
\newtheorem{rem}[theo]{Remark}

\newcommand{\ba}{\begin{array}}
\newcommand{\ea}{\end{array}}
\newcommand{\bea}{\begin{eqnarray}}
\newcommand{\eea}{\end{eqnarray}}
\newcommand{\bead}{\begin{eqnarray*}}
\newcommand{\eead}{\end{eqnarray*}}
\newcommand{\be}{\begin{equation}}
\newcommand{\ee}{\end{equation}}
\newcommand{\bed}{\begin{displaymath}}
\newcommand{\eed}{\end{displaymath}}
\newcommand{\bl}{\begin{lem}}
\newcommand{\el}{\end{lem}}
\newcommand{\bp}{\begin{prop}}
\newcommand{\ep}{\end{prop}}

\newcommand{\bt}{\begin{theo}}
\newcommand{\et}{\end{theo}}
\newcommand{\bc}{\begin{cor}}
\newcommand{\ec}{\end{cor}}
\newcommand{\bd}{\begin{defi}}
\newcommand{\ed}{\end{defi}}

\newcommand{\la}{\Label}
\newcommand{\Label}{\label}
\newcommand{\re}{\Ref}
\newcommand{\Ref}{\ref}

\newenvironment{proof}%
{\begin{sloppypar}\noindent{\bf Proof.}}%
{\hspace*{\fill}$\square$\end{sloppypar}}

\title{Linear
non-autonomous Cauchy problems\\
and evolution semigroups}

\begin{document}

\author{
Hagen Neidhardt\\
WIAS Berlin \\
Mohrenstr. 39\\
10117 Berlin, Germany\\
{\bf E-mail:} neidhard@wias-berlin.de\\
\and
Valentin A. Zagrebnov\\
Universit\'{e} de laM\'{e}diterran\'{e}e
(Aix-Marseille II)\\
Centre de Physique Th\'{e}orique - UMR 6207\\
Luminy - Case 907\\
 13288 Marseille Cedex 9, France\\
{\bf E-mail:} zagrebnov@cpt.univ-mrs.fr
}

\maketitle

\begin{abstract}
The paper is devoted to the problem of existence of propagators for an abstract linear
non-autonomous evolution Cauchy problem of hyperbolic type in separable Banach spaces. The problem
is solved using the so-called evolution semigroup approach which
reduces the existence problem for propagators to a perturbation
problem of semigroup generators. The results are specified to abstract
linear non-autonomous evolution equations in Hilbert spaces where the assumption is made that
the domains of the quadratic forms associated with the generators are
independent of time. Finally, these results are applied to
time-dependent Schr\"odinger operators with moving
point interactions in 1D.\\
\end{abstract}

\noindent
{\bf AMS-subject classification:} Primary 35L90; Secondary  34G10, 47D06\\


\newpage
\tableofcontents

\section{Introduction \rd{and setup of the Problem}}

The aim of the present paper is to develop an approach to Cauchy
problems for linear non-autonomous evolution equations of type
\be\la{1.1}
\frac{\partial}{\partial t} u(t) + A(t)u(t) = 0, \quad u(s) = u_s \in X,
\quad t,s \in {\it I},
\ee
where $\kI$ is a bounded open
interval of $\bR$ and $\{A(t)\}_{t \in \kI}$ is
a family of closed linear operators in the separable Banach space $X$.
Evolution equations of that type are called forward evolution
equations if $s\le t$,
backward if $s \ge t$ and bidirectional evolution equations if $s$ and $t$
are arbitrary.
The main question concerning the Cauchy problem
\eqref{1.1} is to find a so-called ``solution operator'' or
propagator $U(t,s)$ such that $u(t) := U(t,s)u_s$ is in some sense a
solution of \eqref{1.1} satisfying the initial condition
$u(s) = u_s$.

Usually it is assumed that either $\{A(t)\}_{t\in\kI}$ or
$\{-A(t)\}_{t\in\kI}$ are families of generators of $C_0$-semigroups
in $X$. In order to distinct both cases we call \gn{an} operator $A$ a \textit{generator} if it
generates a $C_0$-semigroup \rd{$\{e^{t A}\}_{t\geq0}$}. 
We call $A$  an \textit{anti-generator} if $-A$
generates a
$C_0$-semigroup \rd{$\{e^{- t A}\}_{t\geq0}$, i.e.,  
the operator $-A$ is \gn{the} \textit{generator} of \gn{a} semigroup.}
If simultaneously $A$ is an anti-generator and a
generator, then $A$ is called a group generator.

Very often the Cauchy problem \eqref{1.1} is attacked 
for a suitable dense subset of initial data $u_s$
by solving
it directly in the same manner as ordinary differential equation, \rd{which immediately implies the
existence of
the propagator, see e.g. \cite{Ta1}.} For this purpose one assumes that $\{A(t)\}_{t\in\kI}$
is a family of anti-generators of $C_0$-semigroups such that they uniformly belong
to the class of quasi-bounded semigroups $\kG(M,\gb)$, cf. \cite[Chapter IX]{Ka1}.
If $\{A(t)\}_{t\in\kI}$ is \gn{a} family of anti-generators of class $\kG(M,\gb)$
which are simultaneously anti-generators of \textit{holomorphic} $C_0$-semigroups,
then the evolution equation is called of ``parabolic'' type.
If it is not holomorphic, then it is  called of ``hyperbolic'' type.
In the following in this paper we are \gn{only} interested in the ``hyperbolic'' case.

There is a rich literature on ``hyperbolic'' evolution equations problems.
The first author who discussed these problems was Phillips \cite{Ph1}. A
more general case was considered by Kato in \cite{Ka2,Ka3} and by
Mizohata in \cite{Mi1}. These results were generalized in the sixties in
\cite{El1,Hey1,Ki1,Yo1,Yo2,Kr1,Ha1,Go1}.
Kato has improved these results in two important papers
\cite{Ka6,Ka7}, \rd{where for the first time he
introduced
the assumptions of \textit{stability} and \textit{invariance}.} 
In the seventies and eighties Kato's result were
generalized in  \cite{Do1,Is1,Ko1,Ya1,Ya2}. For related 
results see also \cite{Kom1,Go2,DoFr1}. Recently
several new results were obtained in
\cite{Con1,NiSch1,Ni1,Tan1,Tan2,Tan3}. In the following we refer to these
results as a ``standard approach'' or ``standard methods''. 
Their common feature is that the propagator is
constructed by using certain approximations of the 
family $\{A(t)\}_{t\in\kI}$ for which the corresponding Cauchy
problem can be easily solved. After that one has only to verify 
that the obtained sequence of propagators converges
to the propagator of the original problem. Widely used approximations are a so-called
\textit{Yosida approximation}
introduced in \cite{Yo2}, \textit{piecewise constant} 
approximations proposed by Kato, cf. \cite{Ka6,Ka7},
as well as a combination of both, see \cite{Ki1}.

In contrast to the \textit{standard methods} another approach
was developed in  \cite{Ev1,How,N0,N1,N2,N3}.  It does not rely on any
approximation, \rd{since it is based on the fact that the existence 
problem for the propagator in question is
\textit{equivalent} to \gn{an operator extension problem} for} a suitable
defined operator in a  vector-valued Banach space
$L^p(\kI,X)$  for some $p \in [1,\infty)$. More precisely, it turns
out that any \textit{forward propagator} $\{U(t,s)\}_{(t,s) \in \gD_\kI}$,
$\gD_\kI := \{(t,s) \in \kI \times\kI: s \le t\}$, \gn{(see Definition \ref{II.1a})} 
defines a $C_0$-semigroup in $L^p(\kI,X)$ by
\be\la{1.2ab}
(\kU(\gs) f)(t) := U(t,t-\gs)\chi_{\kI}(t-\gs)f(t-\gs),
\quad f \in L^p(\kI,X),\quad \gs \ge 0,
\ee
where $\chi_{\kI}(\cdot)$ is the characteristic function of the
\gn{open interval} 
$\kI$. $C_0$-semigroups in $L^p(\kI,X)$ admitting a forward propagator
representation \eqref{1.2ab} are called \textit{forward evolution
semigroups}. The anti-generator $K$ of the semigroup $\{\kU(\gs)\}_{\gs\in\bR_+}$, i.e.
$\kU(\gs) = e^{-\gs K}$,
$\gs \in \bR_+$, is called the \textit{forward generator}. 
\rd{Our approach is based on the important
fact that
the set of the forward generators can be described explicitly, 
and that there is a \textit{one-to-one
correspondence} between forward propagators and forward generators , see \cite{N1}.}

\rd{Now, let us assume that the forward propagator $\{U(t,s)\}_{(t,s)\in\gD_\kI}$  is found by the
standard approach
and that it solves the forward evolution equation \eqref{1.1} \gn{in
  some sense}.} Then it turns out that
the forward generator $K_\kI$ defined by \eqref{1.2ab}
is an extension of the so-called \textit{evolution operator} $\widetilde{K}_\kI$ given by
\be\la{1.2aa}
(\widetilde{K}_\kI f)(t) = D_\kI f + Af, \quad f \in
\dom(\widetilde{K}_\kI) = \dom(D_\kI) \cap \dom(A) \ ,
\ee
in $L^p(\kI,X)$ for some $p \in [1,\infty)$, where $D_\kI$ is the
anti-generator of the \textit{right-shift semigroup}
in $L^p(\kI,X)$ and $A$ is the \textit{multiplication} 
operator in $L^p(\kI,X)$ induced by the family
$\{A(t)\}_{t\in\kI}$, see Section 2.

\rd{This remark leads to the main idea of our approach: 
to solve the evolution equation \eqref{1.1} by
\gn{extending} the evolution operator $\widetilde{K}_\kI$ \gn{to}
an anti-generator of \gn{an}  (forward) evolution semigroup. 
Notice that in contrast to the standard
approach now the focus has \gn{moved} from the
\gn{problem to construct} a propagator
to the problem \gn{to find a certain} operator extension.} This so-called  ``extension approach''
or ``extension method'' has a lot of advantages, 
since it works in a very general setting, and it is quite
flexible and transparent. The approach becomes very simple, if the closure
of the evolution operator $\widetilde{K}_\kI$ is already an anti-generator, in other words, if
$\widetilde{K}_\kI$ is  \textit{essentially} anti-generator.
\rd{In this case one gets the forward generator by closing $\widetilde{K}_\kI$, see
Theorem \ref{I.1}, which immediately implies the existence of a unique forward
propagator for the non-autonomous Cauchy problem \eqref{1.1}.}
\rd{Some recent results related \gn{to} the extension
method can be find in e.g. \cite{LM1,MRh1,Ni1,RRhSch1,RSch1}.}

Below we exploit this approach extensively and we show how this method
can be applied to evolution equations of type \eqref{1.1}. We
prove that under the stability and invariance assumptions of Kato
\cite{Ka6,Ka7} the evolution operator $\widetilde{K}_\kI$ 
is already an \textit{essential} anti-generator,
which \gn{means} that its closure $K_\kI$ is a forward generator.

We apply also the extension method to \textit{bidirectional} evolution equations of the type
\be\la{a.1}
i\frac{\partial}{\partial t}u(t) = H(t)u(t),
u(s) = u_s, \quad s,t \in \bR,
\ee
on $\bR$ in Hilbert spaces, where $\{H(t)\}_{t\in \bR}$ is a family of non-negative self-adjoint
operators. Using the extension method we restore and obtain some generalizations of
the Kisy\'{n}ski result \cite{Ki1}. Moreover, we show that 
Kisy\'{n}ski's propagator is in fact the propagator
of an auxiliary evolution equation problem closely related to \eqref{a.1}.
The solution of the auxiliary problem implies a solution for
\eqref{a.1}. The uniqueness of the auxiliary solution
does not imply, however, uniqueness of the original problem \eqref{a.1}, in general.

The paper is organized as follows. \rd{In Section 2
we recall some basic facts of the theory of evolution
semigroups.} Section 3 is devoted to a
perturbation theorem  for generators of these semigroups, which is used then in
Section 4 to show that the closure $K_\kI$ of the evolution operator
\eqref{1.2aa} is an anti-generator.
The results of Section 4 are specified in Section 5 to families $\{A(t)\}_{t\in\bR}$ of
the form $A(t) = iH(t)$ where $H(t)$ are semi-bounded
self-adjoint operators with \textit{time-independent} form domains in a Hilbert
space. In Section 6, we apply these results of Section 5 to 
\gn{Schr\"odinger operators with time-dependent} point interactions of the form:
\bed
H(t) := -\frac{1}{2}\frac{d}{dx}\frac{1}{m(x)}\frac{d}{dx}  + V(x) + \sum^N_{j=1}\gk_j(t)\gd(x-x_j)
\quad x \in \bR, \quad t \in \bR,
\eed
as well as to the case of moving point interactions of the form:
\bed
H(t) := -\frac{d^2}{dx^2} + \gk_1(t)\gd(x - x_1(t)) + \gk_2\gn{(t)}\gd(t)(x - x_2(t))
\eed
where the coupling constants $\gk_j(\cdot)$
are non-negative Lipschitz continuous functions in 
$t \in \bR$ and $x_j(t)$ are $C^2$-trajectories in $\bR$.
\rd{These kind of problems were the subject of publications \cite{D1,DFT1,Po1,SY1,SY2,Y1}.}

\section{Evolution generators}

In the following we are interested not only in the \textit{forward} evolution equations but
also in the \textit{backward} ones as well as in the \textit{bidirectional} evolution equations.
The interest to theses evolution equations  rises from time reversible
problems  in \gn{quantum mechanics},
which we consider in conclusion of this paper as applications. 
For this purpose we show in Section 2.2
how one has to modified the extension approach for backward 
evolution equations. Moreover, in application
to \gn{quantum mechanics} we are concerned with \textit{infinite} time intervals, in
particular, with $\kI = \bR$. In order to apply our approach to this situation it
is useful to \textit{localize} it in time, this means that instead to consider the Cauchy problem on
$\bR$ we consider it on arbitrary finite subintervals of $\bR$.  In
this case, however, one has to  ensure that propagators for different time
intervals are \textit{compatible}.

\subsection{Forward generators}

We start with the definition of a forward propagator in a separable
Banach space.
\bd\la{II.1a}
{\em
Let $X$ be a separable Banach space.
A strongly continuous operator-valued function
$U(\cdot,\cdot): \gD_\kI \longrightarrow \kB(X)$ is called a \textit{forward
propagator} on $\gD_\kI = \{(t,s) \in \kI
\times \kI: s \le t\}$, if

\item[\;\;(i)] $U(t,t) = I_X$ for $t \in \kI$,

\item[\;\;(ii)] $U(t,r)U(r,s) = U(t,s)$ for $(t,r,s) \in \kI^3$, $s \le r \le t$,

\item[\;\;(iii)] $\|U\|_{\kB(X)} := \sup_{(t,s) \in \gD_\kI}\|U(t,s)\|_{\kB(X)} < \infty$.

\rd{\gn{We} call a strongly continuous operator-valued function $U(\cdot,\cdot)$
defined on \gn{$\gD_\bR := \{(t,s) \in \bR \times \bR: s \le t\}$}
a forward propagator, if for any bounded interval $\kI$ the \textit{restriction} of
$U(\cdot,\cdot)$ to $\gD_\kI$ is a forward propagator.}
}
\ed

Another important notion is the  so-called \textit{evolution
operator}. To explain this notion we introduce the Banach space
$L^p(\kI,X)$, $p \in [1,\infty)$, where $X$ is a separable Banach
space. In $L^p(\kI,X)$ we define the \textit{multiplication} operator
\be\la{mult-oper}
(M(\phi)f)(t) := \phi(t)f(t), \quad \dom(M(\phi)) = L^p(\kI,X),
\quad \phi \in L^\infty(\kI).
\ee
\bd
{\em
\rd{A linear operator $K$ in $L^p(\kI,X)$, $p \in
[1,\infty)$,  is called an \textit{evolution operator}, if 
\item[\;\;(i)] it satisfies the conditions:
\be\la{1.12}
\dom(K) \subseteq C(\oI,X),
\ee
\be\la{1.13}
M(\phi)\dom(K) \subseteq \dom(K), \quad \phi \in H^{1,\infty}(\kI),
\ee
and
\be\la{1.14}
K M(\phi)f - M(\phi)K f = M(\dot{\phi})f, \quad f
\in \dom(K), \quad \phi \in  H^{1,\infty}(\kI),
\ee
where $\dot{\phi} := d\phi/dt$, and 
\item[\;\;(ii)] its domain $\dom(K)$ has a dense \textit{cross-section} in
$X$, this means that}
\bed
[\dom(K)]_t := \{x \in X: \exists f \in \dom(K) \; \mbox{such that} \; f(t) = x\},
\eed
is dense in $X$ for each $t \in \kI$. 

If in addition $K$ is an anti-generator or a generator in
$L^p(\kI,X)$, then $K$ is called a \textit{forward} or \textit{backward}
generator, respectively.
}
\ed

\rd{The density of the cross-section is not a trivial condition.} However, one has
to mention that it is important to ensure the \textit{continuity} of the propagator.
Notice that if $K$ is an evolution operator, then its domain $\dom(K)$ is already dense in
$L^p(\kI,X)$, $1 \le p < \infty$.

\rd{Further, by virtue of Theorem 4.12, \cite{N1}, 
it turns out that there is a \textit{one-to-one correspondence}
between the set of forward propagators and the set of forward
generators established by \eqref{1.2ab}. This correspondence plays a
crucial role in our arguments below.}

Let $S_r(\gs)$ be the \rd{right-shift} semigroup in $L^p(\kI,X)$, $1 \le p < +\infty$,
given by
\be\la{1.4a}
(S_r(\gs)f)(t) := f(t-\gs)\chi_{\kI}(t-\gs), \quad f \in L^p(\kI,X).
\ee
This is a $C_0$-semigroup of class $\kG(1,0)$. Its \rd{generator is given by $- D_\kI$}, where
\bed
(D_\kI f)(t) = \frac{\partial}{\partial t}f(t), \quad f \in \dom(D_\kI) :=
H^{1,p}_a(\kI,X), \quad \kI = (a,b).
\eed
\rd{According to our convention \gn{the operator $D_\kI$ is an} anti-generator.} Here
\bed
H^{1,p}_a(\kI,X) := \{f \in H^{1,p}(\kI,X): f(a) = 0\},
\eed
and $H^{1,p}_a(\kI,X)$ is the Sobolev space of $X$-valued absolutely continuous functions
on $\kI$ with $p$-summable derivative.

Notice that a family $\{A(t)\}_{t \in \kI}$ of closed and densely defined linear
operators is called \textit{measurable},
if there is a $z \in \bC$ such that $z$ belongs to the \textit{resolvent
set} $\res(A(t))$ of $A(t)$ for almost every
(a.e.) $t \in \kI$ and for each $x \in X$ the function
\bed
f(t) := (A(t) - z)^{-1}x, \quad t \in \kI,
\eed
is \textit{strongly} measurable. If the family $\{A(t)\}_{t\in\kI}$ is measurable, then
one can show that the multiplication operator $A$,
\be\la{1.5}
(Af)(t) := A(t)f(t), \quad f \in \dom(A),
\ee
\be\la{1.6}
\dom(A)  := \left\{f \in L^p(\kI,X):
\ba{l}
f(t) \in \dom(A(t))\;\mbox{for a.e.} \; t \in \kI,\\
A(t)f(t) \in  L^p(\kI,X)
\ea
\right\}
\ee
is densely defined and closed in $L^p(\kI,X)$.

Instead of solving the Cauchy problem \eqref{1.1} for a suitable set of
initial data $u_s$ we consider the operator
\be\la{1.7}
\widetilde{K}_\kI f := D_\kI f + Af, \quad f \in \dom(\widetilde{K}_\kI) :=
\dom(D_\kI) \cap \dom(A),
\ee
in $L^p(\kI,X)$, $p \in [1,\infty)$. \rd{If \gn{the} domain $\dom({\widetilde{K}}_\kI)$
has a dense cross-section, then \gn{by the definition above}} $\widetilde{K}_\kI$ is an evolution
operator. This leads naturally to following definitions:
\bd\la{I.2}
{\em
Let $\{A(t)\}_{t \in \kI}$ be a measurable family of a closed and
densely defined linear operators in the separable Banach space $X$.

\item[\;\;(i)]
The forward evolution equation \eqref{1.1} is \textit{well-posed} on $\kI$ for some $p \in
[1,\infty)$ if $\widetilde{K}_\kI$ is an evolution
operator.

\item[\;\;(ii)]
\rd{A forward propagator $\{U(t,s)\}_{(t,s) \in \gD_\kI}$ is called a
\textit{solution}} of the well-posed forward evolution equation \eqref{1.1} on
$\kI$ if the corresponding forward generator $K_\kI$, cf. \eqref{1.2ab}, is an
\textit{extension} of $\widetilde{K}_\kI$.

\item[\;\;(iii)] 
\gn{
The well-posed forward evolution equation \eqref{1.1} on
$\kI$ has a unique solution if $\widetilde{K}_\kI$
admits only one extension which is a forward generator.}
}
\ed
It is quite possible that the forward
evolution equation \eqref{1.1} has several solutions, which means that
the evolution operator $\widetilde{K}_\kI$ admits \textit{several} 
extensions, \rd{and each of them is a
forward generator. The dense cross-section property} of the evolution
operator is not sufficient to show that the evolution equation
admits a unique solution.

In the following the next statement will be important for our reasoning.
\bt\la{I.1}
Let $\{A(t)\}_{t \in \kI}$ be a measurable family of closed and
densely defined linear operators in the separable Banach space $X$.
Assume that the forward evolution equation \eqref{1.1} is well-posed on $\kI$
for some $p \in [1,\infty)$. If  the evolution operator
$\widetilde{K}_\kI$ is closable in $L^p(\kI,X)$ and its closure $K_\kI$ is an
anti-generator, then the forward evolution equation \eqref{1.1} on
$\kI$ has a unique solution.
\et
\begin{proof}
Since the evolution equation is well-posed, the domain
$\dom(\widetilde{K}_\kI)$ is densely defined in $L^p(\kI,X)$.
By assumptions the closure $K_\kI$ is an anti-generator. Hence,  it
remains to show that the closure $K_\kI$ satisfies the conditions \eqref{1.12}-\eqref{1.14}.
It is easy to verify that the closure $K_\kI$ satisfies the conditions
\eqref{1.13} and \eqref{1.14}. To show \eqref{1.12} let us assume that $K_\kI$
belongs to $\kG(M,\gb)$. By Lemma 2.16 of \cite{N2} the closure $K_\kI$ admits the estimate
\bed
\|f(t)\|_X \le \frac{M}{(\xi - \gb)^{(p-1)/p}}\|(K_\kI +
\xi)f\|_{L^p(\kI,X)},
\quad f \in \dom(K_\kI), \quad p \in [1,\infty),
\eed
for a.e. $t \in \oI$ and $\xi > \gb$. In particular, we have
\be\la{1.17}
\|f\|_{C(\oI,X)} \le \frac{M}{(\xi - \gb)^{(p-1)/p}}\|(\widetilde{K}_\kI + \xi)f\|_{L^p(\kI,X)},
\quad f \in \dom(\widetilde{K}_\kI).
\ee
Since $\widetilde{K}_\kI$ has a closure $K_\kI$,
there is a sequence of elements $\{f_n\}_{n\in \bN}$ for any $f \in
\dom(\widetilde{K}_\kI)$ such that
$f_n \in \dom(\widetilde{K}_\kI)$, $f_n \longrightarrow f$ and
$\widetilde{K}_\kI f_n \longrightarrow K_\kI f$ in the $L^p(\kI,X)$ sense when
$n \longrightarrow \infty$. By \eqref{1.17} one gets that
$\{f_n\}_{n\in \bN}$ is a Cauchy sequence in
$C(\oI,X)$. Hence $f \in C(\oI,X)$, that proves \eqref{1.12}.
Since $\dom(\widetilde{K}_\kI)$ has a dense cross-section for
each $t \in \kI$,  one gets that its closure $K_\kI$ has a dense cross-section for each
$t \in \kI$. Hence $K_\kI$ is forward generator.

Let $K_\kI$ and $K'_\kI$ be \rd{two different extensions} of $\widetilde{K}_\kI$, which are
both forward generators. Since $K_\kI$ is the closure of
$\widetilde{K}_\kI$ one has $K_\kI  \subseteq K'_\kI$. Since $K_\kI$ and $K'_\kI$
are generators of \rd{a $C_0$-semigroup,} one gets $K_\kI = K'_\kI$.
Hence the evolution equation \eqref{1.1} is uniquely solvable.
\end{proof}

\subsection{Backward generators}

In the following we are also interested in so-called \textit{backward}
evolution equation \eqref{1.1}, $t \le s$, $t,s \in \kI$.
Equations of that type require the introduction of the notion of \textit{backward propagator}:
\bd\la{I.4}
{\em
A strongly continuous  operator-valued function
$V(\cdot,\cdot): \nabla_\kI \rightarrow \kB(X)$
is called a \textit{backward propagator} on 
$\nabla_\kI := \{(t,s) \in \kI \times \kI: t \le s\}$, if

\item[\;\;(i)] $V(t,t) = I_X$ for $t \in \kI$,

\item[\;\;(ii)] $V(t,r)V(r,s) = V(t,s)$ for $(t,r,s) \in \kI^3$, $t \le r \le s$,

\item[\;\;(iii)] $\sup_{(t,s) \in \nabla_\kI}\|V(t,s)\|_{\kB(X)} < \infty$.

\gn{
We call a strongly continuous operator-valued function $V(\cdot,\cdot)$
defined on $\nabla_\bR := \{(t,s) \in \bR \times
\bR: s \le t\}$ a \textit{backward propagator} if for any 
bounded interval $\kI$ the restriction of $V(\cdot,\cdot)$ to
$\nabla_\kI$ is a backward propagator.}
}
\ed
Similar to forward propagators there is a one-to-one correspondence
between backward propagators and backward generators given by
\be\la{1.27}
(e^{\gs K}f)(t) = V(t,t+\gs)\chi_\kI(t+\gs)f(t+\gs),
\quad f \in L^p(\kI,X), \quad \gs \ge 0,
\ee
$p \in [1,\infty)$. With the backward evolution equation we associated the operator
$\widetilde{K}^\kI$
\be\la{1.21}
\widetilde{K}^\kI f = D^\kI f + Af, \quad f \in \dom(\widetilde{K}^\kI) := \dom(D^\kI) \cap \dom(A),
\ee
where
\bed
(D^\kI f)(t) = \frac{\partial}{\partial t}f(t), \quad f\in \dom(D^\kI) :=
\{f \in H^{1,p}_b(\kI,X): f(b) = 0\}
\eed
is the generator of \textit{left-shift} semigroup
$S_l(\gs) = e^{\gs D^\kI}$ on $L^p(\kI,X)$, that is,
\bed
(S_l(\gs)f)(t) = f(t+\gs)\chi_\kI(t+\gs), \quad t \in \kI, \quad
f \in L^p(\kI,X), \quad \gs \ge 0.
\eed
\bd\la{I.3}
{\em
Let $\{A(t)\}_{t \in \kI}$ be a measurable family of closed and
densely defined linear operators in the separable Banach space $X$.

\item[\;\;(i)]
The backward evolution
equation \eqref{1.1} is \textit{well-posed} on $\kI$ for some $p \in [1,\infty)$, if
$\widetilde{K}^\kI$ is an evolution operator.

\item[\;\;(ii)]
A backward propagator $\{V(t,s)\}_{(t,s) \in \nabla_\kI}$ is
called a \textit{solution}  of the well-posed backward evolution equation
\eqref{1.1} on $\kI$ if the corresponding backward generator $K^\kI$, cf. \eqref{1.27}, is an
\textit{extension} of $\widetilde{K}^\kI$.

\item[\;\;(iii)] 
\gn{
The well-posed backward evolution equation \eqref{1.1} on
$\kI$ has a solution if $\widetilde{K}_\kI$
admits only one extension which is a backward generator.}
}
\ed
\rd{Now, following the same line of reasoning as in  Theorem \ref{I.1} we obtain a
similar statement concerning the backward evolution equation \eqref{1.1}:}
\bt\la{I.6}
Let $\{A(t)\}_{t \in \kI}$ be a measurable family of closed and
densely defined linear operators in the separable Banach space $X$.
Assume that the backward evolution equation \eqref{1.1} is well-posed
on $\kI$ for some $p \in [1,\infty)$.
If the evolution operator $\widetilde{K}^\kI$ is closable in
$L^p(\kI,X)$ and its closure $K^\kI$ is a generator, then
the backward evolution equation \eqref{1.1} on $\kI$ has a unique solution.
\et

\subsection{Bidirectional problems}

\gn{Crucial} for studying  \textit{bidirectional} evolution equations on bounded
intervals is the following proposition.
\bp\la{I.7}
Let $\{U(t,s)\}_{(t,s) \in \gD_\kI}$ and $\{V(t,s)\}_{(t,s) \in
\nabla_\kI}$ be for- and backward propagators which correspond to
the for- and backward generators $K_\kI$ and $K^\kI$, respectively. The relation
\be\la{1.30}
V(s,t)U(t,s) = U(t,s)V(s,t) = I_X, \quad (t,s) \in \gD_\kI,
\ee
holds  if and only if
for each $\phi \in H^{1,\infty}_a(\kI) \cap H^{1,\infty}_b(\kI)$ the conditions
\be\la{1.28}
M(\phi)\dom(K_\kI) \subseteq \dom(K^\kI)
\quad \mbox{and} \quad
M(\phi)\dom(K^\kI) \subseteq \dom(K_\kI)
\ee
and
\be\la{1.29}
K^\kI M(\phi)f = K_\kI M(\phi)f,  \quad f \in \dom(K_\kI) \quad \mbox{or} \quad  f
\in \dom(K^\kI),
\ee
are satisfied.
\ep
\begin{proof}
We set
\bed
g(\gs) := e^{\gs K^\kI}M(\phi)e^{-\gs K_\kI}f, \quad f \in L^p(\kI,X), \quad \phi
\in H^{1,\infty}(\kI).
\eed
Taking into account \eqref{1.2ab} and \eqref{1.27} we find
\bed
(g(\gs))(t) =
V(t,t+\gs)\phi(t+\gs)U(t+\gs,t)\chi_\kI(t+\gs)\chi_\kI(t)f(t), \quad t
\in \kI.
\eed
Using \eqref{1.30} we obtain
\be\la{1.29a}
(g(\gs))(t) = \phi(t+\gs)\chi_{(a,b-\gs)}(t)f(t), \quad t \in \kI,
\quad 0 \le \gs < b-a.
\ee
Since
\be\la{1.29b}
(g(\gs)  - M(\phi))f = (e^{\gs K^\kI} - I)M(\phi)f + e^{\gs K^\kI}M(\phi)(e^{-\gs K_\kI} -
I)f ,
\ee
by \eqref{1.29a} we get that
\bed
\lim_{\gs\to+0}\frac{1}{\gs}(g(\gs)  - M(\phi))f = 0, \quad f \in L^p(\kI,X).
\eed
Assuming $f \in \dom(K_\kI)$ we immediately find from \eqref{1.29b} that
$M(\phi)f \in \dom(K^\kI)$ \rd{and \eqref{1.28}}. Interchanging $K^\kI$ and
$K_\kI$ we prove $M(\phi)f \in \dom(K_\kI)$ \rd{and \eqref{1.29}}.

Conversely, assuming \eqref{1.28} and \eqref{1.29} we get that
the function $g(\gs)$ is differentiable and
\bed
\frac{d}{d\gs}g(\gs) = e^{\gs K^\kI}\left(K^\kI M(\phi) - M(\phi)K_\kI\right)e^{-\gs K_\kI}f, \quad \gs \ge 0.
\eed
By virtue of \eqref{1.14} we find
\bed
\frac{d}{d\gs}g(\gs) = e^{\gs K^\kI}M(\dot{\phi})e^{-\gs K_\kI}f, \quad \gs \ge 0
\eed
which yields
\bed
e^{\gs K^\kI}M(\phi)e^{-\gs K}f = M(\phi)f + \int^\gs_0 d\gt\; e^{\gt
  K^\kI}M(\dot{\phi})e^{-\gt K}f, \quad \gs \ge 0.
\eed
Therefore, using representations \eqref{1.2ab} and \eqref{1.27} we obtain
\bead
\lefteqn{
V(t,t+\gs)U(t+\gs,t)\phi(t+\gs)\chi_\kI(t+\gs)f(t) = }
\\
& &
\phi(t)f(t) +
\int^\gs_0 d\gt\; V(t,t+\gt)U(t+\gt,t)\dot{\phi}(t+\gt)\chi_\kI(t+\gt)f(t)
\nonumber
\eead
for $t \in \kI$ and $\gs \ge 0$. Put $s := t + \gs$. Then we get
\bead
\lefteqn{
V(t,s)U(s,t)\phi(s)\chi_\kI(s)f(t) = }
\\
& &
\phi(t)f(t) +
\int^s_t dr\; V(t,r)U(r,t)\dot{\phi}(r)\chi_\kI(r)f(t)
\nonumber
\eead
for $(s,t) \in \gD_\kI$. Let $\overline{\kI}_0 \subset \kI$ be a closed
subinterval such that restriction $\phi\upharpoonright\overline{\kI}_0 = 1$.
If $s,t \in \overline{\kI}_0$, then
\bed
V(t,s)U(s,t)f(t) = f(t)
\eed
for $t \in \overline{\kI}$. Since $[\dom(K_\kI)]_t$ is dense in $X$
for each $t \in \kI$, we prove the first part of \rd{the equality}
\eqref{1.30}. To prove the second part one has to interchange generators $K_\kI$ and
$K^\kI$.
\end{proof}
\bc\la{I.8}
Let $\widetilde{K}_\kI$ and $\widetilde{K}^\kI$, $p \in [1,\infty)$, be
evolution operators in $L^p(\kI,X)$.
Assume that for each $\phi \in H^{1,\infty}_a(\kI) \cap H^{1,\infty}_b(\kI)$ one has
\be\la{1.38}
M(\phi)\dom(\widetilde{K}_\kI) \subseteq \dom(\widetilde{K}^\kI)
\quad \mbox{and} \quad
M(\phi)\dom(\widetilde{K}^\kI) \subseteq \dom(\widetilde{K}_\kI)
\ee
and
\be\la{1.39}
\widetilde{K}^\kI M(\phi)f = \widetilde{K}_\kI M(\phi)f,
\quad f \in \dom(\widetilde{K}_\kI) \quad \mbox{or} \quad  f
\in \dom(\widetilde{K}^\kI).
\ee
If the closures $K_\kI$ and $K^\kI$ of the evolution
operators $\widetilde{K}_\kI$ and $\widetilde{K}^\kI$ exist
and are (respectively) for- and backward generators, then
the corresponding for- and backward propagators
$\{U(t,s)\}_{(t,s) \in \gD_\kI}$ and $\{V(t,s)\}_{(t,s) \in \nabla_\kI}$ verify the
relation \eqref{1.30}.
\ec
\begin{proof}
Let $f \in \dom(\widetilde{K}_\kI)$. Then from \eqref{1.39} and \eqref{1.14} we get
\bed
\widetilde{K}^\kI M(\phi)f = M(\phi)\widetilde{K}_\kI f - M(\dot{\phi})f.
\eed
Since $\widetilde{K}_\kI$ is closable, for each $f \in \dom(K_\kI)$ there is a sequence
$\{f_n\}_{n\in\bN}$, $f_n \in \dim(\widetilde{K}_\kI)$,  such
that $\lim_{n\to\infty}f_n = f$ and
$\lim_{n\to\infty}\widetilde{K}_\kI f_n = K_\kI f$. Since
\bed
\widetilde{K}^\kI M(\phi)f_n = M(\phi)\widetilde{K}_\kI f_n - M(\dot{\phi})f_n, \quad n \in \bN,
\eed
we get $M(\phi)f \in \dom(K^\kI)$ and
\bed
K^\kI M(\phi)f = M(\phi)K_\kI f - M(\dot{\phi})f
\eed
for $f \in \dom(K_\kI)$. Using \eqref{1.14} we prove $M(\phi)\dom(K_\kI)
\subseteq \dom(K^\kI)$ and \eqref{1.29}. Similarly, we prove also
$M(\phi)\dom(K^\kI) \subseteq \dom(K_\kI)$ and \eqref{1.29}. Then application of
Proposition \ref{I.7} completes the proof.
\end{proof}

Now it makes sense to introduce the following definition.
\bd\la{I.9}
{\em
A strongly continuous operator-valued function
$G(\cdot,\cdot): \kI\times\kI \longrightarrow \kB(X)$
is called a \textit{bidirectional propagator} on $\kI \times \kI$ if

\item[\;\;(i)] $G(t,t) = I_X$ for $t \in \kI$,

\item[\;\;(ii)] $G(t,r)G(r,s) = G(t,s)$ for $(t,r,s) \in \kI^3$,

\item[\;\;(iii)] $\sup_{(t,s) \in \kI \times \kI}\|G(t,s)\|_{\kB(X)} < \infty$.

A strongly continuous operator-valued function $G(\cdot,\cdot)$
defined on $\bR \times \bR$ is called a \textit{bidirectional propagator} on $\bR \times \bR$,
if for any bounded interval $\kI$ the restriction of $G(\cdot,\cdot)$ to
$\kI \times \kI$ is a bidirectional propagator.
}
\ed

One can easily verify that if $G(\cdot,\cdot)$ is a bidirectional propagator, then
$U(\cdot,\cdot):= G(\cdot,\cdot)\upharpoonright\gD_\kI$
and $V(\cdot,\cdot):= G(\cdot,\cdot)\upharpoonright\nabla_\kI$
are, respectively, for- and backward propagators related by
\eqref{1.30}. Conversely, if $U(\cdot,\cdot)$ and $V(\cdot,\cdot)$
are, respectively, for- and backward propagators, which are related by \eqref{1.30},
then
\be\la{1.41}
G(t,s) :=
\begin{cases}
U(t,s), & (t,s) \in \gD_\kI\\
V(t,s), & (t,s) \in \nabla_\kI
\end{cases}
\ee
defines a bidirectional propagator.
\bd\la{I.10}
{\em
Let $\{A(t)\}_{t \in \kI}$ be a measurable family of closed and
densely defined linear operators in the separable Banach space $X$.

\item[\;\;(i)]
The evolution equation \eqref{1.1} is \textit{well-posed} on $\kI$ if the for- and backward
evolution equations \eqref{1.1} are well-posed $\kI$ for some $p \in [1,\infty)$.

\item[\;\;(ii)]
The bidirectional propagator $\{G(t,s)\}_{(t,s) \in \kI \times \kI}$
is called a \textit{solution} of the bidirectional evolution equation \eqref{1.1} on
$\kI$ if
the for- and backward propagators $\{U(t,s)\}_{(t,s) \in \gD_\kI}$,
$U(\cdot,\cdot) := G(\cdot,\cdot)\upharpoonright\gD_\kI$, and
$\{V(t,s)\}_{(t,s) \in \nabla_\kI}$, $V(\cdot,\cdot) := G(\cdot,\cdot)\upharpoonright\nabla_\kI$,
are solutions of the for- and backward equations
\eqref{1.1} on $\kI$.

\item[\;\;(iii)]
\gn{
The well-posed evolution equation \eqref{1.1} has a unique solution if
the for- and backward evolution equation \eqref{1.1} has unique solutions.}

}
\ed
\bt\la{I.10a}
Let $\{A(t)\}_{t \in \kI}$ be a measurable family of closed and
densely defined linear operators in the separable Banach space $X$.
Assume that the bidirectional evolution equation \eqref{1.1}
is well-posed on $\kI$ for some $p \in [1,\infty)$. If the closures
$K_\kI$ and $K^\kI$ of evolution operators
$\widetilde{K}_\kI$ and $\widetilde{K}^\kI$ exist in $L^p(\kI,X)$ and are
anti-generators and generators, respectively, then
the bidirectional evolution equation \eqref{1.1} has a unique solution on $\kI$.
\et
\begin{proof}
One easily verifies that the operators $\widetilde{K}_\kI$ and
$\widetilde{K}^\kI$ defined by \eqref{1.7} and \eqref{1.21} satisfy
the conditions \eqref{1.38}, \eqref{1.39}. Then application of Corollary
\ref{I.8} completes the proof.
\end{proof}

\subsection{Problems on $\bR$}

Let us consider the forward evolution equation \eqref{1.1} on $\bR$.
\rd{A natural way to study this problem is to consider the equation
\eqref{1.1} on bounded open intervals $\kI \subset \bR$.} In this case one gets a
solution $\{U_\kI(t,s)\}_{(t,s) \in \bR}$ for any bounded interval
$\kI$. Then we have to guarantee that two
solutions $\{U_{\kI_1}(t,s)\}_{(t,s) \in \gD_{\kI_1}}$ and
$\{U_{\kI_2}(t,s)\}_{(t,s) \in \gD_{\kI_2}}$, which correspond
to different bounded open intervals $\kI_1$ and $\kI_2$, are \textit{compatible},
i.e., one has
\be\la{1.41a}
U_{\kI_1}(t,s) = U_{\kI_2}(t,s), \quad (t,s) \in \gD_1 \subseteq \gD_2,
\ee
for $\kI_1 \subseteq \kI_2$. Below we clarify
this compatibility of propagators in terms of evolution generators.

If $\kI_1 \subseteq \kI_2$, then $L^p(\kI_1,X)$ is a subspace of
$L^p(\kI_2,X)$. Let $Q_{\kI_1}$  denote the
projection from $L^p(\kI_2,X)$ onto the subspace $L^p(\kI_1,X)$
given by
\bed
(Q_{\kI_1}f)(t) := \chi_{\kI_1}(t)f(t), \quad f \in L^p(\kI_2,X).
\eed
Let intervals $\kI_1 = (a_1,b_1)$ and $\kI_2 = (a_2,b_2)$ be related by $a_2 \le
a_1 < b_1 \le b_2$. We set $\kI' = (a_1,b_2)$.
\bp\la{II.9}
Let $\kI_1 = (a_1,b_1)$ and $\kI_2 = (a_2,b_2)$ be two bounded intervals
such that $\kI_1 \subseteq \kI_2$. Further, let $K_{\kI_1}$ and
$K_{\kI_2}$ be forward generators, respectively, in $L^p(\kI_1,X)$ and 
$L^p(\kI_2,X)$. The corresponding propagators
$\{U_{\kI_1}(t,s)\}_{(t,s)\in\gD_{\kI_1}}$
and  $\{U_{\kI_2}(t,s)\}_{(t,s) \in \gD_{\kI_2}}$ are compatible if and only if
for any $f \in L^p(\kI_2,X)$ obeying $Q_{\kI'}f \in \dom(K_{\kI_2})$ one has
$Q_{\kI_1}f \in \dom(K_{\kI_1})$ and relation
\be\la{1.47}
K_{\kI_1}Q_{\kI_1}f = Q_{\kI_1}K_{\kI_2}Q_{\kI'}f.
\ee
\ep
\begin{proof}
We put $K_j := K_{\kI_j}$, $U_j(t,s) :=
U_{\kI_j}$, $Q_j := Q_{\kI_j}$, $j = 1,2$, and $Q' := Q_{\kI'}$.
Assume that the propagators $U_1(t,s)$ and $U_2(t,s)$ are compatible.
In this case one easily verifies that
\bed
e^{-\gs K_1}Q_1f = Q_1e^{-\gs K_2}Q'f, \quad \gs \ge 0, \quad f \in
L^p(\kI_2,X).
\eed
Moreover, by
\bed
\frac{1}{\gs}(I - e^{-\gs K_1})Q_1f = \frac{1}{\gs}Q_1(I - e^{-\gs
  K_2})Q'f, \quad \gs > 0,
\eed
one gets that $Q'f \in \dom(K_2)$ yields $Q_1f \in \dom(K_1)$
as well as \eqref{1.47}.

To prove the converse we set
\bed
W(\gs)f := e^{-(\gt - \gs)K_1}Q_1e^{-\gs K_2 }Q'f,
\quad 0 \le \gs \le \gt.
\eed
If $g := Q'f \in \dom(K_2)$, then $g(\gs) := e^{-\gs K_2}Q'f \in
\dom(K_2)$ for $\gs \ge 0$. Since
\be\la{1.52}
Q'e^{-\gs K_2}Q'f = e^{-\gs K_2}Q'f, \quad f \in L^p(\kI_2,X), \quad
\gs \ge 0,
\ee
we obtain $Q'g(\gs) = Q'e^{-\gs K_1}Q'f \in
\dom(K_2)$, which yields $Q_1e^{-\gs K_2}Q'f
\in \dom(K_1)$. Hence
\bed
\frac{d}{d\gs}W(\gs)f =
e^{-(\gt - \gs)K_1}\left(K_1Q_1 - Q_1K_2\right)e^{-\gs K_2}Q'f,
\quad 0 \le \gs \le \gt.
\eed
Applying to this equation  the relation \eqref{1.47}, we obtain $\partial_{\gs}W(\gs)f = 0$,
which yields
\bed
W(\gt)f = W(0)f, \quad \gt \ge 0,
\eed
or
\be\la{1.52a}
Q_1e^{-\gs K_2}Q'f = e^{-\gs K_1}Q_1f,
\quad \gs \ge 0 ,
\ee
for all those $f \in L^p(\kI_2,X)$ that $Q'f \in \dom(K_2)$.

\rd{Now we notice that the set}
\bed
\kD' := \{f \in L^p(\kI,X): Q'f \in \dom(K_2)\}
\eed
is dense in $L^p(\kI',X)$. Indeed, let $\phi \in H^{1,\infty}(\kI_2)$ such
that $\supp(\phi) \subseteq \overline{\kI'}$. By \eqref{1.13} we have
$M(\phi)f \in \dom(K_2)$ for $f \in \dom(K_2)$. By virtue of
$Q'M(\phi)f = M(\phi)f$, we get $M(\phi)\dom(K_2) \subseteq
\kD_{\kI'}$. Since this holds for any $\phi \in H^{1,\infty}(\kI_2)$
obeying $\supp(\phi) \subseteq \overline{\kI'}$, we immediately find
that $\kD'$ is a dense in $L^p(\kI',X)$.

Hence, the relation \eqref{1.47} holds for any
$f \in L^p(\kI_2,X)$. This implies the compatibility of the
propagators $U_1(t,s)$ and $U_2(t,s)$.
\end{proof}
\bc\la{II.10}
Let $\widetilde{K}_{\kI_1}$ and $\widetilde{K}_{\kI_2}$, $\kI_1 \subseteq \kI_2$, be evolution
operators such that $Q_{\kI'}f \in \dom(\widetilde{K}_{a_2})$ yields $Q_{\kI_1}f \in
\dom(K_{\kI_1})$ and the relation
\be\la{1.52b}
\widetilde{K}_{\kI_1}Q_{\kI_1}f = Q_{\kI_1}\widetilde{K}_{\kI_2}Q_{\kI'}f
\ee
holds. If the closures $K_{\kI_1}$ and $K_{\kI_2}$ of evolution
operators $\widetilde{K}_{\kI_1}$ and $\widetilde{K}_{\kI_2}$ exists
in $L^p(\kI,X)$, $p \in [1,\infty)$, and \rd{they} are forward generators, then the corresponding forward
propagators are compatible.
\ec
\begin{proof}
Let $\kI_1 = (a_1,b_1)$ and $\kI_2 = (a_2,b_2)$, $\kI_1 \subseteq \kI_2$.
As above we set $K_j := K_{\kI_j}$, $U_j(t,s) :=
U_{\kI_j}$, $Q_j := Q_{\kI_j}$, $j = 1,2$, and $Q' := Q_{\kI'}$, where $\kI' = (a_1,b_2)$, see above.
Let $g := Q'f \in \dom(K_{\kI_2})$. Then there is a sequence
$\{g_n\}_{n\in\bN}$, $g_n \in \dom(\widetilde{K}_2)$,
such that $g_n \longrightarrow g$  and $\widetilde{K}_2g_n
\longrightarrow K_2g$ as $n \to \infty$. Let $\phi \in
H^{1,\infty}(\kI_2)$ such that $\supp(\phi) \subseteq \oI'$. By \eqref{1.13} we have
$M(\phi)g_n \in \dom(\widetilde{K}_2)$ and $Q'M(\phi)g_n =
M(\phi)g_n$, $n \in \bN$. Then taking into account \eqref{1.52b} we obtain
\bed
\widetilde{K}_1Q_1M(\phi)g_n =
Q_1\widetilde{K}_2Q'M(\phi)g_n, \quad n \in \bN.
\eed
Using \eqref{1.14} we find
\bed
Q_1\widetilde{K}_2Q'M(\phi)g_n =
Q_2\widetilde{K}_2M(\phi)g_n =
M(\phi)Q_1\widetilde{K}_2g_n + M(\dot{\phi})Q_1g_n,
\quad n \in \bN ,
\nonumber
\eed
which yields
\bed
Q_1\widetilde{K}_2Q'M(\phi)g_n \longrightarrow
Q_1K_2M(\phi)Q'f
\quad \mbox{as} \quad n \to \infty.
\eed
Hence, we obtain
\bed
\widetilde{K}_1Q_1M(\phi)g_n \longrightarrow
Q_1K_2M(\phi)Q'f
\quad \mbox{as} \quad n \to \infty,
\eed
which proves
\bed
\widetilde{K}_1Q_1M(\phi)g_n \longrightarrow
K_1Q_1M(\phi)f
\quad \mbox{as} \quad n \to \infty
\eed
and
\bed
K_1Q_1M(\phi)f = Q_1K_1M(\phi)Q'f.
\eed
Using \eqref{1.14} we also get
\be\la{4.12}
K_1Q_1M(\phi)f = M(\phi)Q_1K_2Q'f +
M(\dot{\phi})Q_1f
\ee
for $\phi \in  H^{1,\infty}(\kI_2)$ obeying $\supp(\phi) \subseteq \oI'$.

Let us put
\bed
\phi_\gd(t) := \left\{
\ba{ll}
0 & t \in (a_2,a_1]\\
{(t-a_1)}/{\gd} & t \in (a_1,a_1 + \gd)\\
1 & t \in [a_1+\gd,b_2)
\ea
\right.
\eed
where $\gd > 0$. Then by  \eqref{4.12} we obtain
\be\la{4.19}
K_1Q_1M(\phi_\gd)f =  M(\phi_\gd)Q_1K_2Q'f +
\frac{1}{\gd}M(\chi_{(a_1,a_1+\gd)})Q_1f
\ee
for any $\gd > 0$. If $g \in L^p(\kI,X)$ is continuous at $t = a_1$ and
$g(a_1) = 0$, then
\bed
s-\lim_{\gt\to 0}\frac{1}{\gd}M(\chi_{(a_1,a_1 + \gd)})g = 0.
\eed
Since $Q'f$ is continuous, one has $f(a_1) = 0$. Hence
\bed
s-\lim_{\gt\to 0}\frac{1}{\gd}M(\chi_{(a_1,a_1 + \gd)})Q'f = 0.
\eed
Since $s-\lim_{\gd\to 0}M(\phi_\gd) = Q'$, from \eqref{4.19} we obtain
that
\bed
\lim_{\gd\to 0}K_1Q_1M(\phi_\gd)f = Q_1K_2Q'f
\eed
and
\bed
\lim_{\gd\to 0}Q_1M(\phi_\gd)f = Q_1f .
\eed
This yields $Q_1f \in \dom(K_1)$ and $K_1Q_1f
= Q_1K_2Q'f$. Applying now Proposition \ref{II.9} one completes the proof.
\end{proof}
%
\bd\la{II.9a}
{\em
Let $\{A(t)\}_{t \in \bR}$ be a measurable family of closed and
densely defined linear operators in the separable Banach space $X$.

\item[\;\;(i)]
The forward evolution equation \eqref{1.1} is \textit{well-posed} on $\bR$ for some $p \in
[1,\infty)$ if for any bounded open interval $\kI$ of $\bR$ the operator
$\widetilde{K}_\kI$ is an evolution operator.

\item[\;\;(ii)]
A forward propagator $\{U(t,s)\}_{(t,s) \in \gD_\bR}$ is called a
\textit{solution} of the well-posed forward evolution equation \eqref{1.1} on
$\bR$ if $\{U_\kI(t,s)\}_{(t,s) \in \gD_\kI}$, $U_\kI(\cdot,\cdot) :=
U(\cdot,\cdot)\upharpoonright\gD_\kI$, is a solution of the  forward
evolution equation \eqref{1.1} for any bounded interval $\kI$ of $\bR$.

\item[\;\;(iii)]
\gn{
The well-posed forward evolution equation \eqref{1.1} on $\bR$ has a unique
solution if for any bounded interval $\kI \subseteq \bR$ the forward evolution
equation \eqref{1.1} admits a unique solution.}

}
\ed
This definition can be extended (\textit{mutatis mutandis}) to backward and
to bidirectional evolution equations on $\bR$.
\bt\la{II.9b}
Let $\{A(t)\}_{t \in \bR}$ be a measurable family of closed and
densely defined linear operators in the separable Banach space $X$.
Assume that the forward evolution equation \eqref{1.1}
is well-posed on $\bR$ for some $p \in [1,\infty)$.
If for any bounded open interval $\kI$ of $\bR$ the closure $K_\kI$ of
the evolution operator $\widetilde{K}_\kI$ exists in $L^p(\kI,X)$, $p \in [1,\infty)$, and it
is an anti-generator, then the forward evolution equation \eqref{1.1} has a unique solution
on $\bR$.
\et
\begin{proof}
Let $\kI_1 \subseteq \kI_2$. One can easily verify that the evolution operators
$\widetilde{K}_{\kI_1}$ and $\widetilde{K}_{\kI_2}$, which  are given
by \eqref{1.7}, satisfy the condition \eqref{1.52b}. Since the operators $\widetilde{K}_{\kI_1}$ and
$\widetilde{K}_{\kI_2}$ are closable and their closures are already forward
evolution generators, one gets from Corollary \ref{II.10} that the
corresponding forward propagators \rd{(they exist and are unique by Theorem
\ref{I.1})} are compatible.
\end{proof}

Proposition \ref{II.9}, Corollary \ref{II.9a} and Theorem \ref{II.9b} can be
generalized (\textit{mutatis mutandis}) to backward and bidirectional evolution equations.

\section{Semigroup \rd{perturbations}}

Theorem \ref{I.1} shows  that the problem of the unique solution  of the
forward or backward evolution
equation \eqref{1.1} can be transformed to the question: whether
the evolution operators $\widetilde{K}_\kI$ or
$\widetilde{K}^\kI$ is are closable and their
closures $K_\kI$ or $K^\kI$ are anti-generators or generators in $L^p(\kI,X)$
for some $p \in [1,\infty)$ ?  In applications $\{A(t)\}_{t\in\kI}$
is often a measurable family of anti-generators or generators belonging
uniformly to  \rd{the class $\kG(M,\gb)$}, for some constants $M$ and $\gb$. One can
easily verify that in this case the induced multiplication operator $A$
is an anti-generator or generator in $L^p(\kI,X)$.

This reduces the problem to the following one: Let $T$ and $A$ be anti-generators or generators in
some Banach space space $\gotX$, is it possible to find conditions ensuring that
their operator sum $\widetilde{K}$:
\be
\widetilde{K}f = Tf + Af, \quad \dom(T)\cap\dom(A),
\ee
is closable in $\gotX$ and its closure $K$ is an anti-generator or generator ?
To prove \rd{such kind of result} we rely on the following theorem.
\bt\la{Ka}
Let in $\gotX$ the operators $T$ and $A$ be generators both belonging  to the class $\kG(1,0)$. If
$\dom(T) \cap \dom(A)$ is dense in $\gotX$ and $\ran(T + A + \xi)$ is
dense in $\gotX$ for some $\xi < 0$, then  $\widetilde{K}$ is
closable and its closure $K$ is a generator from the class $\kG(1,0)$.
\et
This theorem was originally proved by Kato, see
\cite[TheoremIX.2.11]{Ka1}, however, under the additional assumption
that $\widetilde{K}$ is closable. This condition was \gn{dropped} by
Da Prato and Grisvard in \cite[Theorem 5.6]{Gr1}.

In general, the assumption  $T,A \in \kG(1,0)$ is too restrictive for
our purposes. So, we modify this assumption.
It is known that in general it is possible to find in the
Banach space $\gotX$ a new norm such that one of the operators: $T$ or $A$, becomes a generator of the
contraction semigroups on $\gotX$. Indeed, since $T$ is the generator of $C_0$
semigroup, i.e. $T \in \kG(M,\gb_T)$, one has:
\be\la{b.1}
\|e^{\gs T}f\| \le M_Te^{\gb_T\,\gs}.
\ee
Setting
\bed
\bv f \bv  := \sup_{\gs >0}e^{-\gb_T\,\gs}\|e^{\gs \,T}f\|
\eed
one immediately gets that
\bed
\bv e^{\gt T}f \bv = e^{\gb_T\,\gt}
\sup_{\gs > 0}e^{-\gb_T\,\{\gs + \gt\}}\|e^{\gt T}f\| .
\eed
This observation shows that in the Banach space $\gotX$ endowed with
the norm $\bv \cdot\bv$ the semigroup $\{e^{\gs T}\}_{\gs}$ belongs to the class $\kG(1,\gb_T)$
of \textit{quasi-contractive} semigroups. Since
\bed
\|f\| \le \bv f \bv  \le M_T\|f\| ,
\eed
the norm  $\bv \cdot \bv$ is equivalent to $\|\cdot\|$.
The same reasoning can be applied to the semigroup $\{e^{\gs A}\}_{\gs}$, but
in general it is \textit{impossible} to find an equivalent norm
such that \textit{both} semigroups become quasi-contractive.
\bd
{\em
Let $T$ and $A$ be generators of $C_0$-semigroups $e^{\gs T}$ and
$e^{\gs A}$ in $\gotX$. The pair $\{T,A\}$ is called \textit{renormalizable} with
constants $\gb_A$ and $\gb_T$ if
for any sequences $\{\gt_k\}^N_{k=1}$, $\gt_k \ge 0$, and
$\{\gs_k\}^N_{k=1}$, $\gs_k \ge 0$, $n \in \bN$, one has
\be\la{b.2}
\sup_{
\ba{c}
\gt_1 \ge 0, \ldots, \gt_n \ge 0\\
\gs_1 \ge 0, \dots, \gs_n \ge 0\\
n \in \bN
\ea
}
e^{-\gb_T\sum\gt_k}
e^{-\gb_A\sum\gs_k}
\|e^{\gt_1 T}e^{\gs_1 A} \cdots e^{\gt_n T}e^{\gs_n A}f\| < \infty
\ee
for each $f \in \gotX$. In an obvious manner the definition carries
over to
pairs $\{T,A\}$ of anti-generators.
}
\ed
\begin{rem}\la{II.3}
{\em
In the following we formulate the statements in terms of
pairs of generators. However, it is easily to see that these statements
remain true for pairs of anti-generators.
}
\end{rem}
\bl[Lemma 5.1, \cite{N3}]\la{II.1}
Let $T$ and $A$ be generators of $C_0$-semigroups in $\gotX$. There is an
equivalent norm  $\bv\cdot\bv$ on $\gotX$ and such that $T \in
\kG(1,\gb_T)$  and $A \in \kG(1,\gb_A)$
if  and only if the pair $\{T,A\}$ is renormalizable with constants $\gb_T$
and $\gb_A$.
\el
\begin{proof}
Let the pair $\{T,A\}$ be renormalizable with constants $\gb_T$ and
$\gb_A$. \rd{On the space $\gotX$ we define a norm by}
\bed
\bv f\bv :=
\sup_{
\ba{c}
\gt_1 \ge 0, \ldots, \gt_n \ge 0\\
\gs_1 \ge 0, \dots, \gs_n \ge 0\\
n \in \bN
\ea
}
e^{-\gb_T\sum\gt_k}
e^{-\gb_A\sum\gs_k}
\|e^{\gt_1 T}e^{\gs_1 A} \cdots e^{\gt_n T}e^{\gs_n A}f\| \ .
\eed
Obviously, we have $\|f\| \le \bv f\bv$, $f \in
\gotX$. \rd{On the other hand, by the \textit{uniform boundedness principle}, see e.g. \cite[Theorem
I.1.29]{Ka1},} we find that the value of
\bed
M := \sup_{ \ba{c}
\gt_1 \ge 0, \ldots, \gt_n \ge 0\\
\gs_1 \ge 0, \dots, \gs_n \ge 0\\
n \in \bN, \|f\| \le 1 \ea }
e^{-\gb_T\sum\gt_k} e^{-\gb_A\sum\gs_k}
\|e^{\gt_1 T}e^{\gs_1 A} \cdots e^{\gt_n T}e^{\gs_n A}f\|
\eed
is finite, which yields $\bv f\bv \le M\|f\|$, $f \in \gotX$.
Hence, the norms $\|\cdot\|$ and $\bv\cdot\bv$ are equivalent.
Moreover, it turns  out that $T \in \kG(M,\gb_T)$ and $A \in
\kG(M,\gb_A)$. Furthermore, a straightforward computation shows that
\bed
\bv e^{\gt T}f\bv \le e^{\gb_T\gt}\bv f\bv, \quad f \in \gotX,
\eed
\bed
\bv e^{\gs A}f\bv \le e^{\gb_A\gs}\bv f\bv, \quad f \in \gotX.
\eed
Therefore, in the Banach space $\{\gotX,\bv\cdot\bv\}$ the generators
$T$ and $A$ belong, respectively, to $\kG(1,\gb_T)$ and $\kG(1,\gb_A)$.

Conversely, if there is an equivalent norm $\bv\cdot\bv$ in the Banach
space $\gotX$ such that $T \in \kG(1,\gb_T)$ and $A \in \kG(1,\gb_A)$, then a
straightforward computation yields \eqref{b.2}, i.e.,
the pair $\{T,A\}$ is renormalizable with constants $\gb_T$ and $\gb_A$.
\end{proof}
\bd\la{II.4-0}
{\em
Let $\gotY$ be a Banach space which is densely and continuously embedded
into the Banach space $\gotX$, i.e. $\gotY \hookrightarrow \gotX$,  and
let the operator $T$ be  the generator of a
$C_0$-semigroup in $\gotX$. The Banach space $\gotY$ is
called \textit{admissible} with respect to $T$, if the space $\gotY$ is \textit{invariant} with
respect to the semigroup $e^{\gs T}$, i.e.
\bed
e^{\gs T}\gotY \subseteq  \gotY, \quad \gs \ge 0,
\eed
and restriction $e^{\gs \widehat{T}} := e^{\gs T}\upharpoonright \gotY$, $\gs \ge 0$,
\rd{is a $C_0$-semigroup on} $\gotY$.
}
\ed
If $J : \gotY \longrightarrow \gotX$ is the \textit{embedding operator} of
$\gotY$ into $\gotX$, then we get
\bed
e^{\gs T}Jf = Je^{\gs \widehat{T}}f, \quad f \in \gotY,
\eed
which yields
\bed
TJf = J\widehat{T}f, \quad f \in \dom(\widehat{T}).
\eed
\bl\la{II.4}
Let $\widehat{T}$ and $\widehat{A}$ be generators of $C_0$-semigroups
of class $\kG(1,0)$ in the Banach space $\gotY$.
If either $\dom(\widehat{T}^*)$ or $\dom(\widehat{A}^*)$ are dense in $\gotY^*$, then
for any $\xi < 0$ one gets the inequality:
\be\la{b.22}
|\xi|\|g\|_{\gotY^*} \le \|\widehat{T}^*g + \widehat{A}^*g + \xi g\|_{\gotY^*} ,
\quad g \in \dom(\widehat{T}^*) \cap \dom(\widehat{A}^*) .
\ee
\el
\begin{proof}
Let $\dom(\widehat{A}^*)$ be dense in $\gotY^*$. We define
\bed
\widehat{A}_\ga := \widehat{A}(I + \ga \widehat{A})^{-1},
\quad \ga < 0.
\eed
Since $\widehat{A}\in \kG(1,0)$ we have $\widehat{A}_\ga \in \kG(1,0)$ for $\ga <
0$. Further, we set
\bed
\widehat{K}_\ga f := \widehat{T}f + \widehat{A}_\ga f,
\quad f \in \dom(\widehat{K}) := \dom(\widehat{T}), \quad \ga < 0.
\eed
Since $\widehat{T} \in \kG(1,0)$ and $\widehat{A}_\ga \in \kG(1,0)$ 
we find that $\widehat{K}_\ga \in
\kG(1,0)$, $\ga < 0$. This yields the estimate
\bed
\|(\widehat{K}_\ga + \xi)^{-1}f\|_\gotY \le \frac{1}{|\xi|}\|f\|_\gotY,
\quad f \in \gotY, \quad \ga < 0,  \quad \xi < 0.
\eed
Hence, we obtain
\bed
\|(\widehat{K}^*_\ga + \xi)^{-1}g\|_{\gotY^*} \le \frac{1}{|\xi|}\|g\|_{\gotY^*},
\quad g \in \gotY^*, \quad \ga < 0, \quad \xi < 0,
\eed
or
\be\la{b.21}
|\xi|\|g\|_{\gotY^*} \le \|(\widehat{K}^*_\ga + \xi)g\|_{\gotY^*},
\quad g \in \dom(\widehat{K}^*_\ga) = \dom(\widehat{T}^*), \quad \ga < 0, \quad \xi < 0.
\ee
Note that
\bed
\widehat{K}^*_\ga g =
\widehat{T}^*g + \widehat{A}^*_\ga g, \quad g \in \dom(\widehat{T}^*),
\quad \ga < 0, \quad \xi < 0.
\eed
Now, since $\dom(\widehat{A}^*)$ is dense in $\gotY^*$, we get
\bed
s-\lim_{\ga \to 0} (I + \ga \widehat{A}^*)^{-1} = I , \quad \gn{\ga < 0},
\eed
which yields
\bed
\lim_{\ga\to 0}\widehat{K}^*_\ga g = \widehat{T}^*g + \widehat{A}^*g,
\quad \gn{\ga < 0},
\eed
for $g \in \dom(\widehat{T}^*) \cap \dom(\widehat{A}^*)$. \rd{Then in
  the limit \gn{$\ga \to 0$}
the inequality \eqref{b.21} gives  \eqref{b.22}}.

The proof is similar, if one supposes that $\dom(\widehat{T}^*)$ is dense in $\gotY^*$.
\end{proof}
\bc\la{II.5}
Let $T$and $A$ be generators of $C_0$-semigroups of class $\kG(1,0)$
on $\gotX$.
Further, let $\gotY \hookrightarrow \gotX$ be admissible with 
respect to \rd{$T, A$ and \gn{let the} operator
$A$ be such that}
\be\la{b.3}
\gotY \subseteq \dom(A).
\ee
\gn{Assume that} the induced generators $\widehat{T}$ and $\widehat{A}$ \gn{are}
of the class
$\kG(1,0)$. If $\dom(A^*)$ is dense in $\gotX^*$, then
\be\la{b.3a}
|\xi|\|g\|_{\gotY^*} \le
\|\widehat{T}^*g + \widehat{A}^*g + \xi g\|_{\gotY^*},
\quad g \in \dom(\widehat{T}^*) \cap J^*\gotX^*,
 \ee
for $\xi < 0$ where $J: \gotY \longrightarrow \gotX$ is the embedding operator.
\ec
\begin{proof}
By condition \eqref{b.3} we get that $\dom(\widehat{A}^*) \supseteq
J^*\gotX^*$. Let $g \in \dom(\widehat{T}^*) \cap J^*\gotX^*$.
Then there is $h \in \gotX^*$ such that $g = J^*h$. Hence
\bed
\widehat{K}^*_\ga J^*h = \widehat{T}^*J^*h + \widehat{A}^*J^*(I + \ga A^*)^{-1} h, \quad \ga < 0.
\eed
By condition \eqref{b.3} the operator $B := AJ: \gotY \longrightarrow
\gotX$ is bounded. This yields the representation
\bed
\widehat{K}^*_\ga J^*h =  \widehat{T}J^*h + B^*(I + \ga A^*)^{-1} h.
\eed
Since $\dom(A^*)$ is dense in $\gotX^*$ we have $s-\lim_{\ga \to 0}(I
+ \ga A^*)^{-1} = I$. Hence
\bed
\lim_{\ga\to 0} \widehat{K}^*_\ga J^*g = \widehat{T}J^*h + B^*h =
\widehat{T}^*g + \widehat{A}^*g.
\eed
Using \eqref{b.21} we get \eqref{b.3a}.
\end{proof}

\bt[Theorem 5.5, \cite{N3}]\la{II.2}
Let $\{T,A\}$ be a renormalizable pair of generators of
$C_0$-semigroups on $\gotX$. Further, let
the Banach space $\gotY \hookrightarrow \gotX$ be admissible with
respect to operators $T$ and $A$. \gn{Assume that $A$ satisfies condition \eqref{b.3}
and that the pair $\{\widehat{T},\widehat{A}\}$ is renormalizable.}
If either one of the domains $\dom(\widehat{T}^*)$,
$\dom(\widehat{A}^*)$ is dense in $\gotY^*$, or $\dom(A^*)$
is dense in $\gotX^*$, then the closure $K$ of $\widetilde{K}$,
\bed
\widetilde{K}f := Tf + Af, \quad \dom(\widetilde{K}_\kI) = \dom(T) \cap
\dom(A),
\eed
exists and $K$ is  \gn{the} generator of a $C_0$-semigroup.
\et
\begin{proof}
Since the pairs $\{T,A\}$ and $\{\widehat{T},\widehat{A}\}$ are
renormalizable we can assume  without lost of generality
that $T,A \in \kG(1,0)$ as well as  $\widehat{T},\widehat{A}
\in \kG(1,0)$. It is obvious that
\bed
TJf = J\widehat{T}f, \quad f\in \dom(\widehat{T}),
\eed
and
\bed
AJf = J\widehat{A}f, \quad \dom(\widehat{A}).
\eed
By condition \eqref{b.3} we get that
$J^*\gotX^* \subseteq \dom(\widehat{A}^*)$.
Since  $\dom(\widehat{T})$ is dense in $\gotY$ and $\gotY$ is densely
embedded in $\gotX$, we get that the operator
$\widetilde{K}$ is densely defined. In particular, we have
\bed
J \ \dom(\widehat{T}) \subseteq \dom(\widetilde{K}).
\eed
Let $g \in \dom(\widetilde{K}^*) \subseteq \gotX^*$. Then we have
\bed
\left<\widetilde{K}Jf,g\right> = \left<TJf,g\right> +
\left<Bf,g\right> = \left<J\widehat{T}f,g\right> + \left<f,B^*g\right>
\eed
for $f \in \dom(\widehat{T})$. Hence
\bed
\left<\widehat{T}f,J^*g\right> = \left<f,B^*g\right> -
\left<f,J^*\widetilde{K}^*g\right>,
\quad f \in \dom(\widehat{T}),
\eed
which yields $J^*\dom(\widetilde{K}^*) \subseteq \dom(\widehat{T}^*)$.
Since $J^*\gotX^* \subseteq \dom(\widehat{A}^*)$ we obtain
\be\la{b.4}
J^*\dom(\widetilde{K}^*) \subseteq \dom(\widehat{T}^*) \cap
\dom(\widehat{A}^*).
\ee

Now, assume that $\ran(\widetilde{K} + \xi)$ is not dense in $\gotX$ for
some $\xi < 0$. In this case there is a $g \in \gotX^*$ such that
\bed
\left<(\widetilde{K} + \xi)f,g\right> = 0, \quad f \in \dom(\widetilde{K}).
\eed
Hence $g \in \dom(\widetilde{K}^*)$ and $(\widetilde{K}^* + \xi)g = 0$.
By \eqref{b.4} we obtain
\bed
J^*g \in \dom(\widehat{T}^*) \cap \dom(\widehat{A}^*).
\eed
If either $\dom(\widehat{T}^*)$ or $\dom(\widehat{A}^*)$ is dense in
$\gotY^*$, then by Lemma \re{II.4} we get $J^*g = 0$, which yields $g =
0$. If $\dom(A^*)$ is dense in $\gotX^*$, then we apply Corollary
\re{II.5} and find also $J^*g = 0$, which yields $g = 0$. Hence, the range
$\ran(\widetilde{K} + \xi)$ is dense in $\gotX$. We note that by virtue of $T,A \in
\kG(1,0)$  the operator $\widetilde{K}$ is closable. Indeed, one has
the estimate
\bed
|\xi|\|f\| \le \|\widetilde{K}f + \xi f\|, \quad f \in
\dom(\widetilde{K}), \quad \xi < 0,
\eed
which yields the existence of the closure $K$.
Applying now Theorem IX.2.11 of \cite{Ka1} one completes the proof.
\end{proof}
\begin{rem}\la{II.2a}
{\em
Under \gn{the} assumptions of Theorem \ref{II.2} \rd{one can easily verify} that
the set $\gotD := J\dom(\widehat{T}) \subseteq \gotX$ is a core of $K$,
i.e., the closure of the restriction
\gn{$K\upharpoonright\gotD$} coincides with $K$. This
follows from \gn{the} observation that in fact we have proved \gn{the} density of the set
\gn{$(\widetilde{K} + \xi)\gotD$}, $\xi < 0$, in the space $\gotX$.
}
\end{rem}
Taking into account Theorem \ref{II.2} and \cite{Tr1} one immediately obtains the
following corollary.
\bc\la{II.7}
Let the assumptions of Theorem  \ref{II.2} be satisfied. If
either one of the domains $\dom(\widehat{T}^*)$,
$\dom(\widehat{A}^*)$ is dense in
$\gotY^*$, or $\dom(A^*)$ is dense in $\gotX^*$, then the
Trotter product formula
\bed
s-\lim_{n\to\infty} \left(e^{\gs\,T/n}e^{\gs\,A/n}\right)^n = e^{\gs\,K}
\eed
holds uniformly in $\gs \in [0,\gs_0]$, for any $\gs_0 >  0$.
\ec

\section{Solutions of evolution equations}

\subsection{Solutions of forward evolution equations}

Let $\{A(t)\}_{t \in \kI}$ be a measurable family of
anti-generators of class $\kG(M,\gb)$, in the separable Banach space
$X$. By $A$ we denote the multiplication operator induced  by
\eqref{1.5} and \eqref{1.6} in the Banach space
$\gotX = L^p(\kI,X)$, $1 \le p < \infty$.
Notice that $A$ is an anti-generator of a $C_0$-semigroup on $\gotX = L^p(\kI,X)$
of class $\kG(M,\gb)$.
\bd[\cite{Ka6,Ka7}]\la{III.1}
{\em
Let $\{A(t)\}_{t \in \kI}$ be a measurable family of anti-generators of
$C_0$-semigroups in the separable Banach space $X$. The family is called
\textit{forward stable}, if there are constants $M > 0$ and $\gb \ge 0$ such that the estimate
\bed
\|e^{-\gs_1 A(t_1)}e^{-\gs_2 A(t_2)}\cdots e^{-\gs_n A(t_n)}\|_{\kB(X)} \le
Me^{\gb\sum^n_{k=1}\gs_k}
\eed
holds for each sequences $\{\gs_k\}^n_{k=1}$, $\gs_k \ge 0$, and a.e.
$(t_1,t_2,\ldots,t_n) \in \gD_n :=
\{(t_1,t_2,\ldots,t_n) \in \bR^n: a < t_n \le t_{n-1} \le \cdots \le
t_1 < b\}$ with respect of the $\bR^n$-Lebesgue measure.
}
\ed
\rd{It is clear} that if $\{A(t)\}_{t \in \kI}$ is forward stable, then the anti-generators
$A(t)$ belong to $\kG(M,\gb)$ for a.e. $t \in \kI$.
\bl[Lemma 5.9, \cite{N3}]\la{III.2}
Let $\{A(t)\}_{t \in \kI}$ be a measurable family of anti-generators of
$C_0$-semigroups in the separable Banach space
$X$. The pair of anti-generators $\{D_\kI,A\}$ is renormalizable on
$\gotX = L^p(\kI,X)$, $1 \le p < \infty$, if and only if the
family of anti-generators $\{A(t)\}_{t \in \kI}$ is \gn{forward} stable.
\el
\bd\la{III.3}
{\em
Let $\{A(t)\}_{t \in \kI}$ be a measurable family of anti-generators (generators)
of class  $\kG(M,\gb)$ in the separable Banach space
$X$. Further, let  $Y$ be a separable Banach space which is densely and continuously embedded into
$X$. The Banach space $Y$ is called \textit{admissible} with respect to the
family $\{A(t)\}_{t \in \kI}$ if:

\item[\;\;(i)] for a.e. $t \in \kI$, the Banach space $Y$ is
admissible with respect to $A(t)$,

\item[\;\;(ii)] there are constants $\widehat{M}$ and $\widehat{\gb}$
such that the anti-generators (generators) $\{\widehat{A}(t)\}_{t \in \kI}$
of the induced semigroups belong to $\kG(\widehat{M},\widehat{\gb})$
for a.e. $t \in \kI$,

\item[\;\;(iii)] the family  $\{\widehat{A}(t)\}_{t \in \kI}$ is
measurable in $Y$.

}
\ed
We note that the condition (iii) in Definition \ref{III.3} is redundant if $X^*$ is
densely embedded into the Banach space $Y^*$.
\bl[Lemma 5.11, \cite{N3}]\la{III.4}
Let $\{A(t)\}_{t \in \kI}$ be a measurable family of anti-generators
in the separable Banach space $X$ belonging to $\kG(M,\gb)$
and let the separable Banach space $Y$ be densely and continuously embedded into
$X$. The Banach space $\gotY = L^p(\kI,Y)$, $1 \le p < \infty$, is admissible with respect
to the anti-generator $A$ if and only if the family $\{A(t)\}_{t
\in \kI}$ is admissible with respect to $Y$.
\el
Summing up \rd{all those properties it is useful for further purposes} to introduce the following definition:
\bd\la{IV.5a}
{\em
Let $\{A(t)\}_{t \in \kI}$ be a measurable family of
anti-generators in the separable Banach space $X$. Further, let $Y$
be a separable Banach space which is densely and continuously embedded
into $X$. We say the family $\{A(t)\}_{t \in \kI}$ satisfies the
\textit{forward Kato condition} if :

\item[\;\;(i)]  $\{A(t)\}_{t \in \kI}$ is forward stable in $X$,

\item[\;\;(ii)] the Banach space $Y$ is admissible with respect to the
  family  $\{A(t)\}_{t \in \kI}$,

\item[\;\;(iii)] the induced family $\{\widehat{A}(t)\}_{t \in \kI}$ is
  forward stable in $Y$,

\item[\;\;(iv)] $Y \subseteq \dom(A(t))$ holds for a.e. $t \in \kI$,

\item[\;\;(v)] $A(\cdot)\upharpoonright Y \in L^\infty(\kI,B(Y,X))$.

}
\ed
In the following we use a so-called \textit{Radon-Nikodym property} of certain Banach
spaces, see e.g. \cite{DU1}.

We recall that a scalar-valued measure $\mu(\cdot)$ defined on the Borel sets of
$\bR$ satisfies the Radon-Nikodym property if, for instance, its
continuity with respect to the Lebesgue measure implies the existence
of a locally summable function $f(\cdot)$ such that $\mu(\gd) = \int_\gd f(x) dx$
for any bounded Borel set $\gd \subset \bR$. In general, this property \textit{does not}
extend to measures taking their values in Banach spaces. However,
there are classes of Banach spaces where this Radon-Nikodym property
still holds. \rd{For example, \textit{dual} spaces of \textit{separable} Banach spaces admit this
property if and only if they are itself separable.}
This, in particular, yields that the dual Banach space $L^p(\kI,Y)^*$, $1 < p < \infty$,
is isometric to $L^{q}(\kI,Y^*)$, $\frac{1}{p} + \frac{1}{q} = 1$.
\bt\la{III.5}
Let $\{A(t)\}_{t \in \kI}$ be a measurable family of
anti-generators in a separable Banach space $X$. Further, let $Y$
be a separable Banach space which is densely and continuously embedded
into $X$. If $\{A(t)\}_{t \in \kI}$ obeys the forward Kato condition
and if, in addition, one of the following conditions :

\item[{\rm \;\;($A_1$)}]
$Y^*$ satisfies the Radon-Nikodym property,

\item[{\rm \;\;($A_2$)}]
$\dom(\widehat{A}^*(t))$ is dense in $Y^*$ for a.e. $t \in \kI$,

\item[{\rm \;\;($A_3$)}]
$\dom(A(t)^*)$ is dense in $X^*$ for a.e. $t \in \kI$

\item
holds, then the forward evolution equation \eqref{1.1} is well-posed
on $\kI$ for some $p \in (1,\infty)$ and has a unique solution.
\et
\begin{proof}
By Lemma \re{III.2} the pair $\{D_\kI,A\}$ of anti-generators is renormalizable.
Further, let us consider the Banach space $\gotY = L^p(\kI,Y)$, $1
< p < \infty$. Since $Y$ is densely and continuously embedded into
$X$ the Banach space $\gotY$ is densely and continuously embedded in
$\gotX = L^p(\kI,X)$.
Since the family $\{A(t)\}_{t \in \kI}$ is admissible with respect
to $Y$, the operator $A$ is admissible with $\gotY$, cf. Lemma \ref{III.4}. Then
from conditions (iv) and (v) of Definition \ref{IV.5a} we find that $\gotY \subseteq \dom(A)$.

Let (see ($A_1$)) $Y^*$ satisfy the Radon-Nikodym property.
Then $\gotY^* = L^p(\kI,Y)^* = L^q(\kI,Y^*)$, $1/p + 1/q = 1$,
which yields that $\dom(\widehat{D}^*_\kI)$ is dense in $\gotY^*$.
Applying Theorem \ref{II.2} we immediately get that $\widetilde{K}_\kI$
is closable and its closure $K$ generates a $C_0$-semigroup. Taking into account Theorem \ref{I.1}
and Theorem \ref{II.2} we complete the proof of the Theorem under condition ($A_1$).

If $Y$ does not satisfy the Radon-Nikodym property, then the
dual space $\gotY^*$ can be identified with a space
$L^q_w(\kI,Y^*)$, cf. \cite{CM}. The space $L^q_w(\kI,Y^*)$
consists of equivalence classes $[g]$ of $w^*$-measurable functions
$g(\cdot): \kI \longrightarrow Y^*$ such that
$\int^T_0 \|g(t)\|^q_{Y^*}\;dt < \infty$. Two functions
$g_1(\cdot): \kI \longrightarrow Y^*$ and
$g_2(\cdot): \kI \longrightarrow Y^*$ are called \textit{equivalent}, if
$<x,g_1(t)> = <x,g_2(t)>$ holds for a.e. $t \in \kI$ for each $x
\in Y$. Recall that a function $g(\cdot): \kI \longrightarrow Y^*$
is $w^*$-\textit{measurable}, if $<x,g(\cdot)>$ is measurable for each $x \in
Y$. \rd{By a straightforward computation we obtain
that} $(\ga\widehat{A}^* + \xi)^{-1}$, $\xi > \gb$, $\ga > 0$, admits the representation
\be
\left((\ga\widehat{A}^* + \xi)^{-1}g\right)(t) =
(\ga\widehat{A}(t)^* + \xi)^{-1}g(t), \quad g \in L^q_w(\kI,Y^*).
\ee
Hence,
\bed
\left\|\left(\xi(\ga\widehat{A}^* + \xi)^{-1}g - g\right)\right\|^q_{\gotY^*} =
\int^b_a \left\|\xi(\ga\widehat{A}(t)^* + \xi)^{-1}g(t) - g(t)\right\|^q_{Y^*}\;dt.
\eed
Note that for a.e. $t \in \kI$ we have the estimate:
\bed
\left\|\xi(\ga\widehat{A}(t)^* + \xi)^{-1}g(t)\right\|_{Y^*} \le
\frac{\widehat{M}\xi}{\xi - \ga\widehat{\gb}}\;\|g(t)\|_{Y^*} ,
\eed
which yields
\bed
\left\|\xi(\ga\widehat{A}(t)^* + \xi)^{-1}g(t) - g(t)\right\|_{Y^*} \le
\left\{1 + \frac{\widehat{M}\xi}{\xi - \ga\widehat{\gb}}\right\}\|g(t)\|_{Y^*}
\eed
for a.e. $t \in \kI$. Since the domain $\dom(\widehat{A}(t)^*)$ is
dense in $Y^*$ for a.e. $t \in \kI$,  by \textit{assumption} ($A_2$) we get
\bed
\lim_{\ga\to 0}\left\|\xi(\ga\widehat{A}(t)^* + \xi)^{-1}g(t) - g(t)\right\|_{Y^*} = 0
\eed
for a.e. $t \in \kI$.  Hence, by the Lebesgue dominated convergence theorem we obtain
\bed
\lim_{\ga\to 0}\left\|\left(\xi(\ga\widehat{A}^* + \xi)^{-1}g - g\right)\right\| = 0 ,
\eed
which shows that $\dom(\widehat{A}^*)$ is dense in $\gotY^*$.
Taking into account Theorem \ref{I.1} and Theorem \ref{II.2}
\rd{we again conclude that the forward evolution equation (\re{1.1}) 
is well-posed and uniquely solvable.}

Finally, by the same reasoning we obtain that under the assumption ($A_3$)
the domain $\dom(A^*)$ is dense in $\gotX^*$. Applying again Theorem \ref{I.1} and
Theorem \ref{II.2} \rd{we deduce that the evolution equation is well-posed and uniquely solvable.}
\end{proof}

Notice that using \eqref{1.4a} we get the following representation:
\bead
\lefteqn{
\left(\left((e^{-\gs\;D_\kI/n}e^{-\gs\;A/n)}\right)^n f\right)(t) =}\\
& &
e^{-\gs\;A(t - \gs/n)/n}e^{-\gs\;A(t - 2\gs/n)/n}\cdots e^{-\gs\;A(t - \gs)/n}\chi_{\kI}(t-\gs)f(t-\gs)
\nonumber
\eead
for a.e. $t \in \kI$ and $\gs \ge 0$.
\bc\la{IV.6}
If the assumptions of Theorem  \ref{III.5} are satisfied, then
\gn{the propagator can be approximated as follows:}
\bed
\lim_{n\to\infty}\int^{b-\gs}_a
\left\|e^{-\frac{\gs}{n}\,A(s + \frac{n-1}{n}\gs)}e^{-\frac{\gs}{n}\,A(s + \frac{n-2}{n}\gs)}
\cdots e^{-\frac{\gs}{n}\,A(s)}x - U(s + \gs,s)x\right\|^p \;ds = 0
\eed
for each $x \in X$ and $0 \le \gs \le b-a$, \rd{$1 < p <\infty$}.
\ec

\subsection{Backward and bidirectional evolution equations}

To solve the backward evolution equation \eqref{1.1} we assume
that $\{A(t)\}_{t\in\kI}$ is a measurable family of
generators of $C_0$-semigroups of the class $\kG(M,\gb)$. We note that the
multiplication operator defined by \eqref{1.5} and \eqref{1.6}
generates a $C_0$-semigroup of class $\kG(M,\gb)$.
\bd\la{IV.7}
{\em
Let $\{A(t)\}_{t\in\kI}$ be a measurable family of generators of
class $\kG(M,\gb)$ in a separable Banach space
$X$. The family $\{A(t)\}_{t\in\kI}$ is called \textit{backward stable} if
\bed
\|e^{\gs_1A(t_1)}e^{\gs_2A(t_2)} \cdots e^{\gs_nA(t_n)}\|_{\kB(X)} \le
Me^{\gb\sum^n_{k=1}\gs_k}
\eed
is valid for each sequence $\{\gs_k\}^n_{k=1}$, $\gs_k \ge 0$ and
a.e. $t \in \nabla_n := \{(t_1,t_2,\ldots,t_n) \in \bR^n: a < t_1 \le
t_2 \le \ldots \le t_n  < b\}$.
}
\ed
\rd{Then Lemma \ref{III.2} admits the following \gn{analogon}.}
\bl\la{IV.8}
Let $\{A(t)\}_{t \in \kI}$ be a measurable family of semigroup generators
in the separable Banach space $X$, which is supposed to belong to $\kG(M,\gb)$.
Then the pair $\{D^\kI,A\}$ is renormalizable on
$\gotX = L^p(\kI,X)$, $1 \le p < \infty$, if and only if the
family of generators $\{A(t)\}_{t \in \kI}$ is backward stable.
\el
\begin{proof}
Let $\kI = (a,b)$. We introduce the isometry $\mho: L^p(\kI,X)
\longrightarrow   L^p(\kI,X)$, defined by
\be\la{4.1a}
(\mho f)(t) = f(a+b -t), \quad t \in \kI, \quad f \in \dom(\mho) := L^p(\kI,X).
\ee
Notice that $\mho^2 = I$  which yields $\mho^{-1} = \mho$. A straightforward
computation shows that $\mho^{-1}D^\kI \mho = \mho D^\kI \mho = -D_\kI$. Introducing
the family
\bed
A'(t) := -A(a+b-t), \quad t \in \kI,
\eed
and the multiplication operator $A'$ in $L^p(\kI,X)$ we get that
$\mho^{-1}A\mho = \mho A\mho = A'$. Hence, $\mho^{-1}\{D^\kI,A\}\mho = \mho\{D^\kI,A\}\mho =
\{-D_\kI,-A'\}$. Thus, the generator pair $\{D^\kI,A\}$ is renormalizable if and only if
the corresponding anti-generator pair $\{D_\kI,A'\}$ is renormalizable. \rd{From Lemma \ref{III.2} we
obtain} that $\{D_\kI,A'\}$ is renormalizable if and only if the family $\{A'(t)\}_{t\in\kI}$
is forward stable. On the other hand, $\{A'(t)\}_{t\in\kI}$ is forward stable if and only if
$\{A(t)\}_{t\in\kI}$ is backward stable, that finishes the proof.
\end{proof}

\bd\la{IV.5}
{\em
Let $\{A(t)\}_{t \in \kI}$ be a measurable family of
generators on the separable Banach space $X$ and let $Y$
be a separable Banach space which is densely and continuously embedded
into $X$. We say the family $\{A(t)\}_{t \in \kI}$ satisfies the
\textit{backward Kato condition} if:
\begin{enumerate}

\item[(i)]  $\{A(t)\}_{t \in \kI}$ is backward stable in $X$,

\item[(ii)] the Banach space $Y$ is admissible with respect to the
  family  $\{A(t)\}_{t \in \kI}$,

\item[(iii)] the induced family $\{\widehat{A}(t)\}_{t \in \kI}$ (see Definition {\ref{III.3}}) is
backward  stable in $Y$,
\end{enumerate}
and, in addition, we assume that conditions (iv) and (v) of Definition \ref{IV.5a}
are valid.
}
\ed
Then, applying Theorem \ref{II.2} we immediately obtain the following
\rd{statement}:
\bt\la{IV.9}
Let $\{A(t)\}_{t \in \kI}$ be a measurable family of
generators in the separable Banach space $X$. Further, let $Y$
be a separable Banach space which is densely and continuously embedded
into $X$. If $\{A(t)\}_{t \in \kI}$ obeys the backward Kato condition
and if in addition one of the conditions {\rm{(}}$A_1${\rm{)}}-{\rm{(}}$A_3${\rm{)}} holds,
then the backward evolution equation \eqref{1.1} is well-posed on $\kI$ for
some $p \in (1,\infty)$ and has a unique solution.
\et
\bc\la{IV.10}
If the assumptions of Theorem  \ref{IV.9} are satisfied, then for each $x \in X$ we \rd{obtain
approximation of the propagator in the form:}
\bed
\lim_{n\to\infty}\int^{b}_{a +\gs}
\left\|e^{\frac{\gs}{n}\,A(s - \frac{n-1}{n}\gs)}e^{\frac{\gs}{n}\,A(s - \frac{n-2}{n}\gs)}
\cdots e^{\frac{\gs}{n}\,A(s)}x - U(s - \gs,s)x\right\|^p \;ds = 0
\eed
for each $x \in X$ and $0 \le \gs \le b-a$, \rd{$1 < p <\infty$}.
\ec
The proofs \rd{of the Theorem \ref{IV.9} and Corollary \ref{IV.10}} follow directly from Theorem \ref{III.5}
and Corollary \ref{IV.6} by using transformation \eqref{4.1a}.
\bt\la{IV.13}
Let $\{A(t)\}_{t \in \kI}$ be a measurable family of
group generators in the separable Banach space $X$ and let $Y$
be a separable Banach space, which is densely and continuously embedded
into $X$. If  the family $\{A(t)\}_{t \in \kI}$ obeys
the forward and backward Kato conditions and if
one of the conditions {\rm{(}}$A_1${\rm{)}}-{\rm{(}}$A_3${\rm{)}} holds,  then
the bidirectional evolution equation \eqref{1.1} is well-posed on
$\kI$ for some $p \in (1,\infty)$ and has a unique solution.
\et
The proof follows directly from Theorem \ref{I.10a},
Theorem \ref{III.5} and Theorem \ref{IV.9}.
Finally, let us consider bidirectional evolution equations \eqref{1.1}
on $\bR$.
\bt\la{IV.14}
Let $\{A(t)\}_{t \in \bR}$ be a measurable family of
group generators in the separable Banach space $X$.
Further, let $Y$ be a separable Banach space which is densely and continuously embedded
into $X$. If  for any bounded open interval of $\bR$
the family $\{A(t)\}_{t \in \kI}$ obeys
the forward and backward Kato conditions and if
one of the conditions {\rm{(}}$A_1${\rm{)}}-{\rm{(}}$A_3${\rm{)}} holds, then the bidirectional
equation \eqref{1.1} is well-posed on
$\bR$ for some $p \in (1,\infty)$ and admits a unique solution.
\et
The proof follows from a bidirectional modification of Theorem \ref{II.9b}
and from Theorem \ref{IV.13}.

\section{Evolution equations in Hilbert spaces}

Our next aim is to apply the above results to evolution equations for
families of semi-bounded self-adjoint operators $\{H(t)\}_{t\in \bR}$
with \textit{time independent form-domains}.

This case was studied by Kisy\'{n}ski in \cite{Ki1}.
The main Theorem 8.1 of \cite{Ki1} states that if for all elements
of the form-domain, the corresponding closed quadratic form is
continuously differentiable for $t \in \bR$, then one can associated
with the bidirectional evolution equation
\be\la{5.28}
\frac{1}{i} \, \frac{\partial}{\partial t}u(t) + H(t)u(t) = 0,\quad
u(s) = u_s, \quad s,t \in \bR,
\ee
a unique propagator which is called the solution of \eqref{5.28}.
In the present section we elucidate and improve this result.

\subsection{Preliminaries}

Let $\{H(t)\}_{t\in\bR}$ be a family of non-negative
self-adjoint operators in a separable Hilbert space $\gotH$.
In the following we consider the non-autonomous Cauchy problem \eqref{a.1}.
As above we assume that the family of \gn{operators}
$\{H(t)\}_{t\in\bR}$ is measurable. As in \cite{Ki1} we assume also that
\bed
\gotD^+ = \dom(H(t)^{1/2}) \subseteq \gotH, \quad t \in \bR,
\eed
which means that the domain $\dom(H(t)^{1/2})$ is independent of $t
\in \bR$. Introducing the scalar products
\bed
(f,g)^+_t := (\sqrt{H(t)}f,\sqrt{H(t)}g) + (f,g), \quad t \in \bR,
\quad f,g \in \gotD.
\eed
one defines a family of Hilbert spaces $\{\gotH^+_t\}_{t\in \bR}$,
which is densely and continuously embedded, $\gotH^+_t \hookrightarrow \gotH$, into $\gotH$.
The corresponding \rd{vector} norm
is denoted by $\|\cdot\|^+_t$. The natural embedding operator of
$\gotH^+_t$ into $\gotH$ is denoted by $J^+_t: \gotH^+_t \longrightarrow
\gotH$.

By the \textit{closed graph principle} it
follows that for each $t,s \in \bR$ the constants
\bed
c(t,s) := \left\|(H(t) + I)^{1/2}(H(s) + I)^{-1/2}\right\|_{\kB(\gotH)}
\eed
are finite. Obviously, we have
\bed
\|f\|^+_t \le c(t,s)\|f\|^+_s, \quad f \in \gotD, \quad t,s \in \bR,
\eed
which yields the estimates:
\be\la{4.5}
\frac{1}{c(t,s)}\|f\|^+_t \le \|f\|^+_s \le c(s,t)\|f\|^+_{\rd{t}},
\quad f \in \gotD, \quad t,s \in \bR .
\ee
\rd{This means that} the norms $\|\cdot\|^+_t$ are \textit{mutually equivalent}.

We note that for each $t \in \bR$ the Hilbert space $\gotH^+_t$ is
admissible with respect to $H(t)$. The \rd{corresponding \textit{induced group} (see Definition \ref{II.4-0})
is denoted by $U^+_t(\gs)$ and is unitary. Its generator is denoted by $H^+(t)$, i.e.
$U^+_t(\gs) = e^{-i\gs H^+(t)}$. Using the embedding operator $J^+_t$ one gets that
\be\la{4.5a}
U_t(\gs)J^+_tf  = J^+_tU^+_t(\gs)f, \quad f \in \gotH^+_t, \quad \gs \in \bR .
\quad t \in \bR,
\ee
Notice that
\bed
H^+(t)f = H(t)f,
\quad f \in \dom(H^+(t)) := \{f \in \dom(H(t)): H\gn{(t)}f \in \gotH^+_t\} ,
\eed
which gives
\bed
\dom(H^+(t)) = \dom(H(t)^{3/2}).
\eed
The dual space with respect to the \rd{scalar product}  
$(\cdot,\cdot)$ is denoted by $\gotH^-_t$. We note that
\bed
\gotH^+_t \hookrightarrow \gotH \hookrightarrow \gotH^-_t, \quad t
\in \bR.
\eed
The dual space can be obtained as the completion of the Hilbert space $\gotH$ with respect to the norm
\bed
\|f\|^-_t := \|(H(t) + I)^{-1/2}f\|, \quad
f \in \gotH.
\eed
Then from \eqref{4.5} we get
\bed
\frac{1}{c(s,t)}\|f\|^-_t \le \|f\|^-_s \le c(t,s)\|f\|^-_t, \quad f
\in \gotH, \quad t,s \in \bR,
\eed
which shows that the set $\gotD^- := \gotH^-_t$ is
independent of $t$ and the norms $\|\cdot\|^-_t$, $t \in \bR$, are
mutually equivalent. The
natural embedding operator of $\gotH$ into $\gotH^-_t$ is denoted by
$J^-_t: \gotH \longrightarrow \gotH^-_t$. Obviously, we have
\be\la{4.5b}
J^-_t = (J^+_t)^*
\quad \mbox{and} \quad
J^+_t = (J^-_t)^*, \quad t \in \bR.
\ee
The group $U_t(\gs)$, $\gs \in \bR$, $t \in \bR$, admits a unitary extension to the Hilbert space
$\gotH^-_t$, which we denote by $U^-_t(\gs)$, $\gs \in \bR$, $t \in
\bR$. The generator of this group is $H^-_t$,
i.e. $U^-_t(\gs) = e^{-i\gs H^-_t}$, $\gs \in \bR$, $t \in \bR$, and
its domain is given by
\be\la{5.18}
\dom(H^-(t)) = \dom(H(t)^{1/2}) = \gotD^+.
\ee
One can verify that the Hilbert space $\gotH$ is admissible  with
respect to $H^-_t$, $t \in \bR$. The \rd{corresponding unitary} group coincides with
$U_t(\gs)$. One also has
\be\la{4.5c}
U^-_t(\gs)J^-_tf = J^-_t U_t(\gs)f, \quad f \in \gotH,
\quad \gs \in \bR, \quad t \in \bR,
\ee
and
\bed
\dom(H(t)) = \{f \in \dom(H^-(t)): H^-(t)f \in \gotH\}.
\eed
Since $\gotH^+_t$ is admissible with respect to $H(t)$, one gets that
$\gotH^+_t$ is admissible with respect to $H^-_t$. The natural
embedding operator is given by $J_t := J^-_tJ^+_t : \gotH^+_t \longrightarrow \gotH^-_t$,
we obtain:
\bed
U^-_t(\gs)J_tf = J_t U^+_t(\gs)f, \quad f \in \gotH^+_t,
\quad \gs \in \bR, \quad t \in \bR ,
\eed
which shows that
\bed
\dom(H^+(t)) = \{f \in \dom(H^-(t)): H^-(t)f \in \gotH^+_t)\}.
\eed
Moreover, regarding the operator $H(t)$ as an operator \gn{acting}
from $\gotH^+_t$ into $\gotH^-_t$, one finds that $H(t)$ can be extended
to a \textit{contraction} $B(t)$ acting from $\gotH^+_t$ into
$\gotH^-_t$. Indeed, this follows from the estimate
\bea\la{5.19}
\lefteqn{ \|B(t)f\|^-_t  = }\\
& & \|(H(t) + I)^{-1/2}H(t)f\|_t \le \|H(t)^{1/2}f\|_t \le \|f\|^+_t,
\quad f \in \dom(H(t)).
\nonumber
\eea
Finally, taking into account \eqref{4.5a}-\eqref{4.5c} we get the relations:
\bed
U^+_t(\gs)^* = U^-_t(-\gs)
\quad \mbox{and} \quad
U^-_t(\gs)^* = U^+_t(-\gs),
\quad \gs \in \bR, \quad t \in \bR.
\eed

\subsection{Auxiliary evolution equation}

We consider the Hilbert space
\bed
X := \gotH^-_{\rd{t=0}}   \quad \mbox{with} \quad \|\cdot\|_X := \|\cdot\|^-_{\rd{t=0}}
\eed
and the auxiliary bidirectional evolution equation
\be\la{5.21}
\frac{\partial}{\partial t}u(t) + iH^-(t)u(t) = 0
\ee
on $\bR$. To apply results from Section 4 we set $A(t) = iH^-(t)$, $t \in \bR$.
Obviously, $\{A(t)\}_{t\in\bR}$ is a family of group generators
in $X$. Further, we set
\be\la{5.12}
Y := \gotH^+_{\rd{t=0}}  \quad \mbox{with} \quad \|\cdot\|_Y := \|\cdot\|^+_{\rd{t=0}} .
\ee
It turns out that the Hilbert space $Y = \gotH^+_0$ is densely and
continuous embedded into $X$ and admissible
with respect to $\{A(t)\}_{t\in\bR}$.
\bl\la{V.1}
Let $\{H(t)\}_{t\in\bR}$ be a measurable family of non-negative
self-adjoint operators defined in a separable Hilbert space
$\gotH$ such that $\dom(H(t)^{1/2})$ is independent of $t \in
\bR$. If $\kI$ is a bounded open interval and
\bed
c_\kI := \sup_{(t,s) \in \kI\times\kI}c(t,s) < \infty,
\eed
then there are constants $M_\kI$ and $\gb_\kI$ such that
$\{A(t)\}_{t\in\kI}$ is a measurable
family of group generators belonging to $\kG(M_\kI,\gb_\kI)$.

If the Hilbert space $Y$ is given by \eqref{5.12} and there is a
constant $\gga_\kI > 0$ such that
\be\la{5.23}
c(t,s) \le e^{\gga_\kI|t-s|}, \quad t,s \in \kI,
\ee
holds, then the families $\{A(t)\}_{t\in\kI}$ obey the forward and backward Kato
conditions, respectively.
\el
\begin{proof}
The measurability of the family $\{A(t)\}_{t \in \kI}$ follows from
the equivalence of weak and strong measurability, see e.g. \cite{HiPhi}.
Next, we have
\bead
\lefteqn{
\|e^{\gs A(t)}x\|_X = \|e^{i\gs H^-(t)}x\|^-_0 \le
c(0,t)\|e^{i\gs H^-(t)}x\|^-_t  }\\
& &
\le c(0,t)\|x\|^-_t \le
c(0,t)c(t,0)\|x\|^-_0 = c(0,t)c(t,0)\|x\|_X,
\nonumber
\eead
$\gs \in \bR$. Hence,
\bed
\|e^{\gs A(t)}x\|_X \le M_\kI\|x\|_X, \quad x \in X, \quad \gs \in \bR,
\quad t \in \kI,
\eed
where $M_\kI := c^2_\kI$, which yields that $A(t)$ generates a group of the class $\kG(M_\kI,0)$.

If condition \eqref{5.23} is satisfied, then the forward and backward stability of
$\{A(t)\}_{t\in\kI}$ follows from  \cite[Theorem 4.3.2]{Ta1}.

To prove the measurability of $\{A(t)\}_{t\in\kI}$ we note that $Y$
is admissible for a.e. $t \in \kI$. Using \eqref{4.5} we obtain that
the generator $\widehat{A}(t)$ of the induced group \rd{(Definition \ref{II.4-0})} belongs to
$\kG(M_\kI,0)$, too. The measurability of the induced family $\{\widehat{A}(t)\}_{t\in \kI}$
follows from the equivalence of strong and weak measurability.

The forward and backward stability of $\{\widehat{A}(t)\}_{t\in \kI}$ follows again from
condition \eqref{5.23} and \cite[Theorem 4.3.2]{Ta1}.

The condition $Y \subseteq \dom(A(t))$ for a.e. $t \in \kI$ is
obtained  from \eqref{5.18}. \rd{The \gn{condition}
$A(\cdot)\upharpoonright Y \in L^\infty(\kI,\kB(Y,X))$
follows from \eqref{5.19}.}
\end{proof}
\bt\la{V.2}
Let $\{H(t)\}_{t\in\bR}$ be a measurable family of non-negative
self-ad\-joint operators defined in a separable Hilbert space
$\gotH$ such that \gn{the} domain $\dom(H(t)^{1/2})$ is independent of $t \in
\bR$. If for any bounded open interval $\kI$ the condition
\eqref{5.23} is satisfied, then the auxiliary  bidirectional
evolution \rd{problem} \eqref{5.21} is well-posed on $\bR$ for $p \in (1,\infty)$ and has a unique solution
$\{G^-(t,s)\}_{(t,s)\in\bR\times\bR}$ obeying the estimate
\be\la{5.23a}
\|G^-(t,s)x\|^-_t \le e^{\gga_\kI(t-s)}\|x\|^-_s, \quad x \in \gotH^-_s,
\ee
for all $(t,s) \in \kI \times \kI$.
\et
\begin{proof}
Since $Y = \gotH^+_0$ is a Hilbert space, all conditions
($A_1$)-($A_3$) are satisfied.
Using Lemma \ref{V.1} and Theorem \eqref{IV.14}
one gets that the bidirectional evolution equation \eqref{5.21}
has a unique solution $\{G^-(t,s)\}_{(t,s) \in \bR \times \bR}$ on $\bR$.

By Corollary \ref{IV.6} there is a subsequence $\{n_k\}_{k \in \bN}$
such that one has
\bed
U^-_\kI(s+\gs,s)x = s-\lim_{k\to\infty}
e^{-i\frac{\gs}{n}H^-(s + \frac{n_k-1}{n_k}\gs)}
e^{-i\frac{\gs}{n}H^-(s + \frac{n_k-2}{n_k}\gs)}
\cdots
e^{-i\frac{\gs}{n}H^-(s)}x
\eed
for each $x \in \gotH^-_s$ and a.e. $s \in (a, b-\gs)$, $0 \le \gs \le
b-a$,  where $U^-_\kI(\cdot,\cdot) :=
G^-(\cdot,\cdot)\upharpoonright\gD_\kI$.
This yields the estimate
\bed
\|U^-_\kI(s+\gs,t)x\|^-_{s + \gs} \le
e^{\gga_\kI\gs}\|x\|^-_s, \quad x \in \gotH^-_s,
\eed
for a.e. $s \in (a,b-\gs)$, $0 \le \gs \le b-a$. Since $U^-_\kI(\cdot,\cdot)$ is strongly
continuous, this holds for any $s \in (a,b-\gs)$.  Setting $t := s +
\gs$ we obtain
\be\la{5.23b}
\|U^-_\kI(t,s)x\|^-_t \le
e^{\gga_\kI |t-s|}\|x\|^-_s, \quad x \in \gotH^-_t, \quad (t,s) \in \gD_\kI.
\ee
Similarly, using Corollary \ref{IV.10} we obtain
\bed
\|V^-_\kI(s - \gs,t)x\|^-_{s - \gs} \le
e^{\gga\gs}\|x\|^-_s, \quad x \in \gotH^-_s,
\eed
for $s \in (a+\gs,b)$, $0 \ge \gs \ge b-a$, where $V^-_\kI(\cdot,\cdot) :=
G^-(\cdot,\cdot) \upharpoonright \nabla_\kI$.  Hence one gets the inequality:
\be\la{5.23c}
\|V^-_\kI(t,s)x\|^-_t \le
e^{\gga_\kI |t-s|}\|x\|^-_s, \quad x \in \gotH^-_t, \quad (t,s) \in \nabla_\kI.
\ee
Using \eqref{5.23b} and \eqref{5.23c} we immediately obtain
\rd{\eqref{5.23a}}.
\end{proof}

\subsection{Back to the original problem}

Our Theorem \ref{V.2} gives no information about \textit{solvability} of the bidirectional
evolution equation \eqref{a.1} on $\bR$.
This goes back to the fact that in general the evolution equation
might be \textit{not} well-posed. In fact, it may happen that the
cross-sections of the sets
\bed
\dom(\widetilde{K}_\kI) := \dom(D_\kI) \cap \dom(H_\kI) =
H^{1,p}_a(\kI,\gotH) \cap \dom(H_\kI)
\eed
and
\bed
\dom(\widetilde{K}^\kI) = \dom(D^\kI) \cap \dom(H_\kI) =
H^{1,p}_b(\kI,\gotH) \cap \dom(H_\kI),
\eed
$p \in (1,\infty)$,
are \textit{not dense} in $\gotH$ for intervals $\kI = (a,b) \subseteq \bR$.
Recall that \gn{$H_\kI$ is defined as the multiplication operator} induced by the family
$\{H(t)\}_{t\in\kI}$ in $L^p(\kI,\gotH)$.

To avoid this situation we assume in the following that the
bidirectional evolution \rd{problem} \eqref{5.28} is well-posed on $\bR$.
Naturally, then we face up to the question: whether under this condition
the evolution equation \eqref{5.28} admits a solution on $\bR$?
\bl\la{V.3}
Let $\{H(t)\}_{t\in\bR}$ be a measurable family of non-negative
self-adjoint operators defined in the separable Hilbert space
$\gotH$ such that $\dom(H(t)^{1/2})$ is independent of $t \in
\bR$. 
If for any bounded open interval $\kI$ the condition
\eqref{5.23} is satisfied, then there is a unitary bidirectional propagator
$\{G(t,s)\}_{(t,s) \in \bR^2}$ on $\gotH$, such that
\be\la{5.28a}
J^-_0G(t,s) = G^-(t,s)J^-_0, \quad (t,s) \in \bR^2.
\ee
Moreover, there is a bidirectional propagator
$\{G^+(t,s)\}_{(t,s) \in \bR^2}$ on $\gotH^+_0$, such that
\be\la{5.28b}
J_0G^+(t,s) = G^-(t,s)J_0, \quad (t,s) \in \bR^2
\ee
and
\be\la{5.28c}
J^+_0G^+(t,s) = G(t,s)J^+_0, \quad (t,s) \in \bR^2.
\ee
\el
\begin{proof}
\gn{Let $J^+ := J^+_0$, $J^- := J^-_0$ and $J := J_0$}.
We consider the forward case.
Let $\kI = (a,b)$ be a bounded open interval of $\bR$ and let $0 \le
\gs \le b-a$. By Corollary \ref{IV.6} we get that 
\bead
\lefteqn{
U^-(\cdot+\gs,\cdot)J^-x_0 =}\\
& &
s\stackrel{L^p(\kI_\gs,X)}{\longrightarrow}\lim_{n\to\infty}
e^{-i\frac{\gs}{n}H^-(\cdot + \frac{n-1}{n}\gs)}e^{-i\frac{\gs}{n}H^-(\cdot + \frac{n-2}{n}\gs)}
\cdots e^{-i\frac{\gs}{n}H^-(\cdot)}J^-x_0,
\nonumber
\eead
$\kI_\gs := (a,b-\gs)$, for each $x_0 \in \gotH$. Since
\bead
\lefteqn{
e^{-i\frac{\gs}{n}H^-(s + \frac{n-1}{n}\gs)}e^{-i\frac{\gs}{n}H^-(s + \frac{n-2}{n}\gs)}
\cdots e^{-i\frac{\gs}{n}H^-(s)}J^-x_0 =}\\
& &
J^-e^{-i\frac{\gs}{n}H(s + \frac{n-1}{n}\gs)}e^{-i\frac{\gs}{n}H(s + \frac{n-2}{n}\gs)}
\cdots e^{-i\frac{\gs}{n}H(s)}x_0
\nonumber
\eead
for a.e. $s \in \kI_\gs$  and since
$\{e^{-i\frac{\gs}{n}H(\cdot + \frac{n-1}{n}\gs)}e^{-i\frac{\gs}{n}H(\cdot + \frac{n-2}{n}\gs)}
\cdots e^{-i\frac{\gs}{n}H(\cdot)}\}_{n\in\bN}$ is bounded in
$L^2(\kI_\gs,\gotH)$, we obtain that the \rd{weak} limit
\bed
U(\cdot + \gs,\cdot)x_0 := w\stackrel{L^p(\kI_\gs,\gotH)}{\longrightarrow}\lim_{n\to\infty}
e^{-i\frac{\gs}{n}H(\cdot + \frac{n-1}{n}\gs)}e^{-i\frac{\gs}{n}H(\cdot + \frac{n-2}{n}\gs)}
\cdots e^{-i\frac{\gs}{n}H(\cdot)}x_0
\eed
exists for each $x_0 \in \gotH$ and for each $\gs \in (0,b-a)$. Hence,  we
obtain
\bed
J^-U(s+\gs,s)x_0 = U^-(s+\gs,s)J^-x_0
\eed
for a.e. $s \in \kI_\gs$, $\gs \in (0,b-a)$ and any $x_0 \in \gotH$. We
note that
\bed
\|U(s+\gs,s)x_0\|_\gotH \le \|x_0\|_\gotH
\eed
for a.e. $s \in \kI_\gs$ and $\gs \in (0,b-a)$, $x_0 \in
\gotH$. Taking into account that the propagator
$\{U^-(t,s)\}_{(t,s)\in\gD_\kI}$ is strongly continuous, one gets that
$\{U(t,s)_{(t,s)\in\gD_\kI}$ is a weakly continuous family
of contractions obeying
\be\la{5.34}
J^-U(t,s)x_0 = U^-(t,s)J^-x_0
\ee
for any $(t,s)\in \gD_\kI$ and for each $x_0 \in \gotH$.
Similarly one proves that there is a weakly
continuous family of contractions $\{V(t,s)\}_{(t,s) \in \nabla_\kI}$
such that
\be\la{5.35}
J^-V(t,s)x_0 = V^-(t,s)J^-x_0
\ee
holds for $(t,s)\in \nabla_\kI$ and $x_0 \in \gotH$.
Setting $G(t,s) := U(t,s)$, $(t,s) \in \gD_\kI$, and
$G(t,s) := V(t,s)$, $(t,s) \in \nabla_\kI$, and taking into account that $\kI$ is arbitrary, we
obtain a weakly continuous family $\{G(t,s)\}_{(t,s)\in \bR\times\bR}$ of contractions
obeying
\bed
G(t,s) = G(s,t)^{-1}, \quad (t,s) \in \bR \times \bR.
\eed
Since for $(t,s) \in \bR \times \bR$ and any $x_0 \in \gotH$ one has
\bed
\|x_0\|_\gotH = \|G(s,t)G(t,s)x_0\|_\gotH \le
\|G(t,s)x_0\|_\gotH \le \|x_0\|_\gotH ,
\eed
$\|G(t,s)x_0\|_\gotH = \|x_0\|_\gotH$, which shows that $\{G(t,s)\}_{(t,s) \in \bR \times \bR}$
is a weakly continuous family of unitary operators. However, this immediately
yields that $\{G(t,s)\}_{(t,s) \in \bR \times \bR}$ is in fact a strongly
continuous family of unitary operators obeying
\be\la{5.35a}
J^-G(t,s) = G^-(t,s)J^-, \quad (t,s) \in \bR \times \bR ,
\ee
which yields that $\{G(t,s)\}_{(t,s) \in \bR\times\bR}$ is a unitary
propagator.

Now we put
\bed
V^+(s,t) := U^-(t,s)^*, \;\; (t,s) \in \gD_\bR,
\quad \mbox{and} \quad
U^+(s,t) := V^-(t,s)^*, \;\; (t,s) \in \nabla_\bR,
\eed
as well as
\bed
G^+(s,t) := G^-(t,s)^*, \quad (t,s) \in \bR^2.
\eed
Then one can easily verify that $\{G^+(t,s)\}_{(t,s) \in \kI\times\kI}$ is
weakly continuous propagator for any bounded interval $\kI$. Taking
into account \eqref{5.23a} and \eqref{5.23b} we obtain
\bed
\|V^+(s,t)y\|^+_s \le e^{\gga(t-s)}\|y\|^+_t, \quad y \in \gotH^+_t,
\eed
and
\bed
\|U^+(t,s)y\|^+_t \le e^{\gga(t-s)}\|y\|^+_s, \;\; y \in \gotH^+_s,
\eed
for $s \le t$. Using the scalar product $(f,g)^+_s :=
(\sqrt{H(s) + I}f,\sqrt{H(s) + I}g)$, $f,g \in \gotD^+$, we get:
\bed
(\|U^+(t,s)y - y\|^+_s)^2 = (\|U^+(t,s)y\|^+_s)^2 + (\|y\|^+_s)^2 -
2\real(U^+(t,s)y,y)^+_s.
\eed
Now, using \eqref{5.23a} we find
\bed
\|U^+(t,s)y\|^+_s \le e^{\gga(t-s)}\|U^+(t,s)y\|^+_t \le
e^{2\gga(t-s)}\|y\|^+_s,
\eed
which implies
\bed
(\|U^+(t,s)y - y\|^+_s)^2 \le e^{4\gga(t-s)}(\|y\|^+_s)^2 +
(\|y\|^+_s)^2 - 2\real(U^+(t,s)y,y)^+_s.
\eed
By the weak continuity of the forward propagator $\{U^+(t,s)\}_{(t,s)
  \in \gD_\bR}$ we obtain $\lim_{t\to s+0}U^+(t,s) = I$. Hence,
$\lim_{t\to s+0}\|U^+(t,s)y - y\|^+_s = 0$ for each $y \in \gotH^+_s$.
Since the norms $\|\cdot\|^+_t$ and $\|\cdot\|^+_0$ are equivalent, we
find $\lim_{t\to s+0}\|U^+(t,s)y - y\|^+_0 = 0$ for each $y \in Y = \gotH^+_0$.
Similarly we prove $\lim_{t\to s-0}\|V^+(t,s)y - y\|^+_0 = 0$ for each
$y \in Y = \gotH^+_0$. Using the representation:
\bed
G^+(t,s) = G^+(t,0)G^+(0,s) \ ,
\eed
where
\bed
G^+(t,0) =
\begin{cases}
U^+(t,0), & t \ge 0,\\
V^+(t,0), & t \le 0,
\end{cases}
\quad \mbox{and} \quad
G^+(0,s) =
\begin{cases}
V^+(0,s), & s \ge 0,\\
U^+(0,s), & s \le 0,
\end{cases}
\eed
one proves the strong continuity of  the families: $\{G^+(t,0)\}_{(t\in\bR}$ and
$\{G^+(0,s)\}_{s \in \bR}$, which yields the strong continuity of $\{G^+(t,s)\}_{(t,s)\in\bR^2}$.

Finally, by $(J^-)^* = J^+$ and $J = J^-J^+$ we find the equation
\bed
J^+G^+(s,t) = G(s,t)J^+, \quad (s,t) \in \bR \times \bR,
\eed
which by virtue of \eqref{5.35a} proves \eqref{5.28c}. Hence we get that
\bead
\lefteqn{
JG^+(s,t) = J^-J^+G^+(s,t) = }\\
& &
J^-G(s,t)J^+ =G^-(s,t)J^-J^+ = G^-(s,t)J,
\quad \quad (s,t) \in \bR \times \bR,
\eead
which proves \eqref{5.28b}.
\end{proof}

\rd{Now it is useful to introduce} the following definition.
\bd\la{V.3a}
{\em
Let $\{G(t,s)\}_{(t,s)\in \bR \times \bR}$ be a
bidirectional propagator in a separable Banach space $X$ and
let $Y$ be a separable Banach space, which is densely and continuously embedded into $X$.
The Banach space $Y$ is called admissible with respect to the family
$\{G(t,s)\}_{(t,s)\in\bR\times\bR}$  if there is
a bidirectional propagator $\{\widehat{G}(t,s)\}_{(t,s)\in \bR \times
  \bR}$  in $Y$ such that
\be\la{5.35b}
G(t,s)J = J\widehat{G}(t,s), \quad (t,s) \in \bR \times \bR,
\ee
holds where $J$ is the embedding operator of $Y$ into $X$.
}
\ed
The following theorem generalizes Theorem 8.1 of \cite{Ki1}. Our proof
is quite independent from the that in \cite{Ki1}.
\bt\la{V.4}
Let $\{H(t)\}_{t\in\bR}$ be a measurable family of non-negative
self-ad\-joint operators defined in the separable Hilbert space
$\gotH$ such that $\dom(H(t)^{1/2})$ is independent of $t \in
\bR$. If the bidirectional evolution equation \eqref{5.28} is well-posed on $\bR$
for some $p \in (1,\infty)$ and the condition \eqref{5.23} is satisfied for any
bounded open interval, then the bidirectional evolution equation
\eqref{5.28} admits on $\bR$ a unitary solution $\{G(t,s)\}_{(t,s)\in \bR \times \bR}$
for which the Hilbert space $\gotH^+_0$ is admissible.
Moreover, if for any bounded open interval $\kI = (a,b)$ the
sets
\bea\la{5.35c}
H^{1,p}_a(\kI,\gotH^+_0) \cap \dom(H_\kI)
\quad \mbox{and} \quad
H^{1,p}_b(\kI,\gotH^+_0) \cap \dom(H_\kI), \quad p \in (1,\infty),
\eea
are dense in $H^{1,p}_a(\kI,\gotH^+_0)$ and
$H^{1,p}_b(\kI,\gotH^+_0)$, respectively, then there is only one
unitary solution for which the Hilbert space $\gotH^+_0$ is admissible.
\et
\begin{proof}
We have to show that the evolution operator $\widetilde{K}_\kI$,
\bed
\widetilde{K}_\kI f = D_\kI + iH_\kI f, \quad f \in \dom(\widetilde{K}_\kI) =
\dom(D_\kI) \cap \dom(H_\kI),
\eed
which is associated with the forward evolution equation \eqref{5.28},
can be extended to a forward generator.
Let $\widetilde{K}^-_\kI$ be evolution operator:
\bed
\widetilde{K}^-_\kI g = D^-_\kI g + iH^-_\kI g, \quad g \in \dom(\widetilde{K}^-_\kI)
= \dom(D^-_\kI) \cap \dom(H^-_\kI),
\eed
associated with \eqref{5.21}, where $D^-_\kI$ is the anti-generator of the
right-shift semigroup in $L^p(\kI,\gotH^-_0)$, and let $H^-_\kI$ be
multiplication operator induced by $\{H^-(t)\}_{t\in\kI}$.
By $\kJ^-$ we denote the embedding operator of $L^p(\kI,\gotH)$,
$p \in (1,\infty)$, into $L^p(\kI,\gotH^-_0)$, defined as:
\bed
(\kJ^-f)(t) = J^-f(t), \quad f \in L^p(\kI,\gotH)
\eed
where $J^- := J^-_0$. One can easily verify that $\kJ^-\dom(\widetilde{K}_\kI) \subseteq
\dom(\widetilde{K}^-_\kI)$ and
\bed
\widetilde{K}^-_\kI\kJ^-f = \kJ^-\widetilde{K}_\kI f, \quad f\in
\dom(\widetilde{K}_\kI).
\eed
By Theorem \ref{V.2} the forward evolution equation \eqref{5.21} is
uniquely solvable. This means that the  operator $\widetilde{K}^-_\kI$ admits only one
extension $K^-_\kI$, which is a forward generator. In fact, it
has been already proven that the closure of $\widetilde{K}^-_\kI$ coincides with
$K^-_\kI$.

By Lemma \ref{V.3} there is a forward generator $\{U_\kI(t,s)\}_{(t,s) \in
\gD_\kI}$, $U_\kI(t,s) := G(t,s)\upharpoonright\gD_\kI$ obeying \eqref{5.28a}. By
the relation
\bed
(e^{-\gs K_\kI}f)(t) = U_\kI(t,t-\gs)\chi_\kI(t-\gs)f(t-\gs), \quad f \in
L^p(\kI,\gotH),
\eed
one defines a forward generator $K_\kI$ in $L^p(\kI,\gotH)$.
Obviously, we have
\bed
e^{-\gs K^-_\kI}\kJ^- f = \kJ^- e^{-\gs K_\kI}f, \quad f \in L^p(\kI,\gotH).
\eed
Hence
\bed
\kJ^-\dom(K_\kI) \subseteq \dom(K^-_\kI)
\eed
and
\bed
K^-_\kI\kJ^-f = \kJ^-K_\kI f, \qquad f \in \dom(K_\kI).
\eed
Notice that
\bed
e^{-\gs K^-_\kI}g = g - \int^\gs_0 d\gt\;e^{-\gt K^-_\kI}K^-_\kI g, \quad g
\in L^p(\kI,\gotH^-).
\eed
Then choosing $g = \kJ^-f$, $f \in \dom(\widetilde{K}^-_\kI)$, we obtain
\bed
\kJ^-e^{-\gs K_\kI}f = \kJ^-f - \kJ^-\int^\gs_0 d\gt \;e^{-\gt
  K_\kI}\widetilde{K}_\kI f
\eed
which yields
\bed
e^{-\gs K_\kI}f = f - \int^\gs_0 d\gt \;e^{-\gt K_\kI}\widetilde{K}_\kI f,
\quad f \in \dom(\widetilde{K}_\kI).
\eed
Therefore,  $\widetilde{K}_\kI \subseteq K_\kI$, which shows that
$\{U_\kI(t,s)\}_{(t,s)\in \gD_\kI}$ is a solution of the forward
evolution equation \eqref{5.28} on $\kI$. The same procedure can be
applied to the backward evolution equation \eqref{5.28} on $\kI$. Hence
the unitary bidirectional propagator $\{G(t,s)\}_{(t,s) \in \bT
  \times \bR}$ defined by \eqref{5.28a} is,
in fact, a solution of the bidirectional evolution equation \eqref{5.28} on $\bR$.

Assume now that $\{Z(t,s)\}_{(t,s)\in \bR \times \bR}$ is another
unitary solution of the bidirectional evolution equation \eqref{5.28}
such that Hilbert space $\gotH^+_0$ is admissible with respect to
$\{Z(t,s)\}_{(t,s)\in \bR \times \bR}$. Then from
\bed
J^+\widehat{Z}(t,s)  = Z(t,s)J^+, \quad (t,s) \in \bR \times \bR,
\eed
we obtain
\bed
\widehat{Z}(t,s)^*J^- = J^-Z(t,s)^*, \quad (t,s) \in \bR \times \bR ,
\eed
where it is used that $J^- = (J^+)^*$.
We set $Z^-(t,s) := \widehat{Z}(s,t)^*$, $(t,s) \in \bR \times \bR$.
Since $\{Z(t,s)\}_{(t,s)\in\bR\times\bR}$ is unitary, we have $Z(t,s) =
Z(s,t)^*$. By this we find
\bed
Z^-(t,s)J^- = J^-Z(t,s), \quad (t,s) \in \bR \times \bR.
\eed
Since $\{Z(t,s)\}_{(t,s) \in \bR \times \bR}$ and
$\{Z^+(t,s)\}_{(t,s)\in\bR\times\bR}$ are
bidirectional propagators in $\gotH$ and $\gotH^+_0$, respectively,
one easily gets that $\{Z^-(t,s)\}_{(t,s)\in\bR\times\bR}$ is a
bidirectional propagator in
$\gotH^-_0$. For any bounded interval $\kI$ in $\bR$
a forward generator $L^-_\kI$ corresponds to the forward
propagator $\{Z^-_\kI(t,s)\}_{(t,s)\in\gD_\kI}$,
$Z^-_\kI(\cdot,\cdot) := Z^-(\cdot,\cdot) \upharpoonright \kI\times\kI$
by relation:
\bed
(e^{-\gs L^-_\kI}f)(t) := Z^-_\kI(t,t-\gs)\chi_\kI(t-\gs)f(t-\gs),
\quad t \in \kI, \quad f \in L^P(\kI,\gotH^-_0).
\eed
It is obvious that
\bed
e^{-\gs L^-_\kI}\kJ^- = \kJ^-e^{-\gs L_\kI}, \quad \gs \ge 0,
\eed
where $L_\kI$ denotes the forward generator, which
corresponds to $\{Z_\kI(t,s)\}_{(t,s)\in \gD_\kI}$, $Z_\kI(t\cdot,\cdot)
:= Z(\cdot,\cdot)\upharpoonright\gD_\kI$. Hence,  $\kJ^-\dom(L_\kI)
\subseteq \dom(L^-_\kI)$ and
\bed
L^-_\kI\kJ^-f = \kJ^-L_\kI f, \quad f \in \dom(K_\kI).
\eed
Since $L_\kI$ is an extension of $\widetilde{K}_\kI$, we obtain
\bed
L^-_\kI\kJ^-f = \kJ^-\widetilde{K}_\kI f, \quad f \in \dom(\widetilde{K}_\kI),
\eed
which shows that $L^-_\kI$ is an extension of
$\widetilde{L}^-_\kI :=
L^-_\kI\upharpoonright\kJ^-\dom(\widetilde{K}_\kI)$.
Since
\bed
K^-_\kI\kJ^-f = \kJ^-\widetilde{K}_\kI f, \quad f \in \dom(\widetilde{K}_\kI).
\eed
holds one gets that $K^-_\kI$ is also an extension of
$\widetilde{L}^-_\kI$. Since the intersection
$H^{1,p}_a(\kI,\gotH^+_0) \cap \dom(H_\kI)$, cf. \eqref{5.35c},
is dense in $H^{1,p}_a(\kI,\gotH^+_0)$, the domain $\dom(\bar{L}^-_\kI)$
of the closure $\bar{L}^-_\kI$ of 
$\widetilde{L}^-_\kI$ contains $H^{1,p}_a(\kI,\gotH^+)$.
By Remark \ref{II.2a} the set $H^{1,p}_a(\kI,\gotH^+_0)$ is a
\textit{core} of $K^-_\kI$, which shows that $K^-_\kI = \bar{L}^-_\kI$.
Hence $L^-_\kI = K^-_\kI$, which yields $Z^-_\kI(t,s) = U^-_\kI(t,s)$,
$(t,s) \in \gD_\kI$, for any bounded interval $\kI$ of $\bR$. The same
can be proven for the backward evolution equation, which ensures that
the bidirectional evolution \eqref{5.28} admits only \textit{one} solution for which the
Hilbert space $\gotH^+_0$ is admissible.
\end{proof}

\section{Examples}

\subsection{Point interactions with varying coupling constant}

We consider a family
$\{H(t)\}_{t\in \bR}$ of self-adjoint operators associated in the Hilbert space $\gotH = L^2(\bR)$ 
with the sesquilinear forms
\bea\la{6.50a}
\lefteqn{
\goth_t[f,g] := }\\
& &
\int_\bR \left\{\frac{1}{2m(x)}f'(x)\overline{(g'(x)}\right\}
 + V(x)f(x)\overline{g(x)} +
\sum^N_{j=1} \gk_j(t)f(x_j)\overline{g(x_j)},
\nonumber
\eea
where $f,g \in \dom(\goth_t) := H^{1,2}(\bR)$, $1 \le N \le \infty$. We assume that
\bed
m(x) > 0, \quad \frac{1}{m} + m \in L^\infty(\bR),
\quad \mbox{and} \quad V \in L^\infty(\bR)
\eed
$x_j \in \bR$, $j = 1,2,\dots,N$, and that the coupling constants $\gk_j(\cdot): \bR
\longrightarrow \bR_+$ are measurable functions.
The family $\{H(t)\}_{t\in\bR}$ is uniformly semibounded from
below. Indeed, we have
\bed
H(t) \ge -\|V\|_{L^\infty(\bR)}, \quad t \in \bR.
\eed
Therefore, without loss of generality we assume that $V(x) \ge 0$ for
a.e. $x \in \bR$, which yields that $\{H(t)\}_{t\in\bR}$ is a family of
non-negative self-adjoint operators. Moreover, one can easily verify
that $\{H(t)\}_{t \in \bR}$ is a measurable family of self-adjoint operators.
\rd{For finite $N$ \gn{the} domain $\dom(H(t))$ admits an explicit description. Indeed, in this
case the operators $H(t)$ are given by the sum of operators in the form-sense (\ref{6.50a}):}
\bed
H(t) = -\frac{1}{2}\frac{d}{dx}\frac{1}{m(x)}\frac{d}{dx} \dotplus V(x) \dotplus
\sum^N_{j=1}\gk_j(t)\gd(x-x_j)
\eed
with domain defined by 
\bea\la{6.5}
\lefteqn{
\dom(H(t)) :=}\\
& &
\left\{f \in H^{1,2}(\bR):
\ba{l}
\tfrac{1}{m}f' \in H^{1,2}(\bR \setminus \bigcup^N_{j=1}\{x_j\}),\\
\left(\tfrac{1}{2m}f'\right)(x_j - 0) - \left(\tfrac{1}{2m}f'\right)(x_j + 0) = \gk_j(t)f(x_j),\\
j = 1,2,\ldots,N < \infty
\ea
\right\}
\nonumber
\eea
for $t \in \kI$. In the following we assume \rd{(\textit{convergence} condition)} that
\be\la{6.6}
\sup_{t\in\kI}\sum^N_{j=1}\gk_j(t) < \infty, \quad 1 \le N \le \infty,
\ee
for each bounded subinterval $\kI \subset \bR$. Furthermore, we assume \rd{(\textit{continuity} condition)} 
that for each bounded subinterval $\kI \subset \bR$ there is a
constant $C_\kI > 0$ such that
\be\la{6.7}
\sum^N_{j=1} |\gk_j(t) - \gk_j(s)| \le C_\kI|t-s|, \quad t,s \in \kI.
\ee

Since $\gotD^+ := \dom(H(t)^{1/2}) =
\dom(\goth_t) = H^{1,2}(\bR)$  is independent of $t\in\bR$,
\rd{Theorem \ref{V.2} is applicable in this case: the \textit{auxiliary}
bidirectional evolution equation \eqref{5.21} admits a unique solution,
if the estimate \eqref{5.23} is satisfied for each bounded subinterval $\kI
\subset \bR$.} 

To show this it is sufficient to verify that the estimate
\be\la{6.8}
\|\sqrt{H(t) + I}f\| \le e^{\gga_\kI |t-s|}\|\sqrt{H(s) + I}f\|, \quad f,g
\in \gotD^+ 
\ee
holds for any $t,s \in  \kI$. Indeed, one obviously has
\bed
|f(x_j)|^2 = 2\real\left\{\int^{x_j}_{-\infty} f'(x)\overline{f(x)} dx\right\},
\quad f \in H^{1,2}(\bR), \quad j \in 1,2,\ldots,N,
\eed
which yields
\be\la{6.10}
|f(x_j)|^2 \le \int_\bR \left\{|f'(x)|^2 + |f(x)|^2\right\} dx \ , \ 
j = 1,2,\ldots,N.
\ee
Hence
\be\la{6.10a}
|f(x_j)|^2 \le \max\{1,2\|m\|_{L^\infty}\}\|\sqrt{H(s) + I}f\|^2,
\quad j = 1,2,\ldots,N.
\ee
Therefore, we have
\bed
\left|\|\sqrt{H(t) + I}f\|^2 - \|\sqrt{H(s) + I}f\|^2\right|
\le
\sum^N_{j=1}|\gk_j(t) - \gk_j(s)|\;|f(x_j)|^2.
\eed
and consequently, by \eqref{6.10} we obtain:
\bead
\lefteqn{
\left|\|\sqrt{H(t) + I}f\|^2 - \|\sqrt{H(s) + I}f\|^2\right|
\le }\\
& &
\max\{1,2\|m\|_{L^\infty}\}\|\sqrt{H(s) + I}f\|^2\sum^N_{j=1}|\gk_j(t) - \gk_j(s)|.
\nonumber
\eead
Using \eqref{6.7} we get
\be\la{6.14}
\left|\|\sqrt{H(t) + I}f\|^2 - \|\sqrt{H(s) + I}f\|^2\right| \le
2\gga_\kI\;|t-s|\;\|\sqrt{H(s) + I}f\|^2
\ee
for $t,s \in \kI$,  where
\bed
\gga_\kI := \frac{1}{2}C_\kI \max\{1,2\|m\|_{L^\infty}\}.
\eed
From \eqref{6.14} it follows that
\bed
\|\sqrt{H(t) + I}f\| \le \sqrt{1 + 2\gga_\kI\;|t-s|}\;\|\sqrt{H(s) + I}f\| \ ,
\eed
which yields
\bed
\|\sqrt{H(t) + I}f\| \le (1 + \gga_\kI\;|t-s|)\;\|\sqrt{H(s) + I}f\| \,,
\eed
for $t,s \in \kI$. Since $1 + \gga_\kI\;|t-s| \le e^{\gga_\kI |t-s|}$,
for any $t,s \in \kI$, we obtain \eqref{6.8}.

Then by Theorem \ref{V.4} the \textit{original} bidirectional evolution equation
\eqref{5.28} admits a solution for which the Hilbert space
$H^{1,2}(\bR)$ is admissible. It is more complicated to solve the
problem whether this solution of the \textit{original} problem is \textit{unique}.
To this end one has to verify the \rd{additional condition (\ref{5.35c})} of Theorem
\ref{V.4}. This condition is satisfied if the sets
$(I + H_\kI)^{-1}H^{1,2}_a(\kI,\gotH)$  and $(I + H_\kI)^{-1}H^{1,2}_b(\kI,\gotH)$
are dense in $H^{1,2}_a(\kI,\gotH^+_0)$ and $H^{1,2}_b(\kI,\gotH^+_0)$,
for any bounded interval $\kI = (a,b)$, respectively.

To prove this we introduce linear operators  $C_j: L^2(\bR) \longrightarrow \bC$ defined by
\bed
C_jf := ((I + H(0))^{-1/2}f)(x_j) \,, \qquad f \in L^2(\bR)\,, \qquad j =
1,2,\ldots, N \ .
\eed
Using the estimate \eqref{6.10a} we find $|C_jf| \le
C\|f\|_{L^2(\bR)}$, where $C$ is given by $C := \max\{1,2\|m\|_{L^\infty}\}$.
Setting $B_j := C^*_jC_j$ we obtain the representation
\bed
(I + H(t))^{-1} = (I + H(0))^{-1/2}R(t)
(I + H(0))^{-1/2} \ , 
\qquad  t \in \bR \,,
\eed
where
\bed
R(t) := \left(I +\sum^N_{j=1}\gk_j(t)B_j\right)^{-1}, \qquad t \in \bR.
\eed
Since the coupling constants are locally Lipschitz continuous,
see \eqref{6.7}, we get that $R(t)x \in H^{1,2}_a(\kI,\gotH)$, $x \in \gotH$, for
any bounded open interval $\kI \subseteq \bR$. Hence, $R(t)f(t) \in
H^{1,2}_a(\kI,\gotH)$ for $f \in H^{1,2}_a(\kI,\gotH)$ and
any bounded open interval $\kI \subseteq \bR$. Hence we get
$(I + H(t))^{-1}f(t) \in H^{1,2}_a(\kI,\gotH^+_0)$ for $f \in H^{1,2}_a(\kI,\gotH)$ and
$\kI \subseteq \bR$. Now we show that the set of
elements $(I + H(t))^{-1}f(t)$, $f \in
H^{1,2}_a(\kI,\gotH)$, is \textit{dense} in $H^{1,2}_a(\kI,\gotH^+_0)$.
Note that the standard norm of $H^{1,2}_a(\kI,\gotH^+_0)$ is
equivalent to the norm
\bed
\|f\|_{H^{1,2}_a(\kI,\gotH^+_0)} = \left(\int_\kI \|\sqrt{I + H(0)}f'(t)\|^2_\gotH\;dt\right)^{1/2} .
\eed
If the elements $(I + H(t))^{-1}f(t)$, $f \in
H^{1,2}_a(\kI,\gotH)$, are not dense in $H^{1,2}_a(\kI,\gotH^+_0)$,
then there is an element $g \in H^{1,2}_a(\kI,\gotH^+_0)$ such that
{\small
\bed
\int_\kI
\left(\sqrt{I + H(0)}
\left((I + H(0))^{-1/2}R(t)(I + H(0))^{-1/2}f(t)\right)',\sqrt{I + H(0)}g'(t)\right) dt = 0
\eed
}
for any $f \in H^{1,2}_a(\kI,\gotH)$. Hence we obtain
\bed
\int_\kI (R'(t)(I + H(0))^{-1/2}f(t) + R(t)(I + H(0))^{-1/2}f'(t),\sqrt{I + H(0)}g'(t)) dt = 0\,.
\eed
Setting $h(t) := (I + H(0))^{-1/2}f(t) \in H^{1,2}_a(\kI,\gotH^+_0)$
and $k(t) := \sqrt{I + H(0)}g'(t) \in L^2(\kI,\gotH)$ we find that
\be\la{6.14a}
\int_\kI (R'(t)h(t) + R(t)h'(t),k(t)) dt = 0
\ee
for any $h \in H^{1,2}_a(\kI,\gotH^+_0)$. Since
$H^{1,2}_a(\kI,\gotH^+_0)$ is dense in $H^{1,2}_a(\kI,\gotH)$ one gets
that \eqref{6.14a} holds for any $h \in H^{1,2}_a(\kI,\gotH)$. From
\eqref{6.14a} we obtain
\bed
\int_\kI (h'(t), R(t)k(t)) dt = -\int_\kI (h(t),R'(t)k(t)) dt
\eed
for any $h \in H^{1,2}_a(\kI,\gotH)$, which yields $z(t) := R(t)k(t) \in
H^{1,2}_b(\kI,\gotH)$ and
\be\la{6.14b}
\frac{d}{dt}R(t)k(t) - R'(t)k(t) = 0
\ee
for a.e. $t \in \kI$. From the representation
\bed
k(t) = \left(I + \sum^N_{j=1}\gk_j(t)B_j\right)z(t)
\eed
and condition \eqref{6.7} we obtain that $k(t) \in
H^{1,2}_b(\kI,\gotH)$.
Taking into account this \rd{last} observation we get from \eqref{6.14b} that
$R(t)k'(t) = 0$ for a.e $t \in \kI$. Since $\ker(R(t)) = \{0\}$ for $t
\in \kI$,  we find that $k'(t) = 0$, which implies $k(t) = const$. But since $k(b)
= 0$, we get $k(t) = 0$ for $t \in \kI$. Hence $g'(t) = 0$ for
$t \in \kI$, which yields $g(t) = 0$ for $t \in \kI$. Consequently, the
set $(I + H_\kI)^{-1}H^{1,2}_a(\kI,\gotH)$ is dense in
$H^{1,2}_a(\kI,\gotH^+_0)$ for any bounded open interval $\kI = (a,b)$.

Similarly, one proves that the set $(I + H_\kI)^{-1}H^{1,2}_b(\kI,\gotH)$
is dense in $H^{1,2}_b(\kI,\gotH^+_0)$ for any bounded open interval $\kI = (a,b)$.

Taking into account the second part of Theorem
\ref{V.4} one finds that there is a \textit{unique} solution of the
original  problem \eqref{5.28} such that $\gotH^+_0$ is admissible.

Therefore, summing up this line of reasoning we obtain the proof of the following theorem:
\bt\la{VI.1}
Let $0 \le V \in L^\infty(\bR)$, $m > 0$ and ${1}/{m} + m \in
L^\infty(\bR)$. Further, let $\{x_j\}_{j\in\bN}$ be  a 
\rd{{\rm{(}}infinite{\rm{)}}} sequence of real
numbers which are mutually different and let $\gk_j(\cdot): \bR \longrightarrow \bR_+$
be non-negative locally Lipschitz continuous functions. Moreover, let
$\{H(t)\}_{t\in\bR}$ be a family of non-negative self-adjoint
Schr\"odinger operators associated with the sesquilinear forms
\eqref{6.50a}. If the conditions \eqref{6.6} and \eqref{6.7} are
satisfied, then the bidirectional
evolution equation \eqref{5.28} is well-posed on $\bR$ for $p = 2$ and 
possesses a unique solution
$\{G(t,s)\}_{(t,s)\in\bR\times\bR}$ such that 
$H^{1,2}(\bR)$ is admissible.
\et
\rd{A similar problem was treated in three
dimensions by \cite{SY2} for the case of finite point interactions and $m(x) = const$. 
In contrast to Theorem \ref{VI.1} their results concern the case of coupling constants 
$\gk_j(t)$ which  are \textit{twice continuously differentiable}, cf. \cite[Theorem 1]{SY2}.} 
In this case the bidirectional evolution equation is verified in the strong sense. Moreover, 
only the existence of a bidirectional propagator was established under the weaker assumption
that the coupling constants $\gk_j(t)$ are locally $L^\infty$-function, cf. \cite[Theorem 2]{SY2}. 
The first results was improved in \cite{DFT2}, where the
smoothness of the coupling constants was reduced to a certain H\"older
continuity. However, it seems to be difficult to extend the technique
used \cite{DFT2,SY2} to the  case of an
\textit{infinite} number of point interactions and to a \textit{non-smooth} position
dependent effective mass $m$.

In conclusion we would like to remark that Theorem \ref{VI.1} covers rather 
\textit{bizarre} situations. For instance, let $\{x_j\}_{j \in \bN}$ be an enumeration of the rational 
numbers $\bQ$ and let $\{\gk_j(t)\}$ be a sequence of coupling constants such
that conditions \eqref{6.6} and \eqref{6.7} are satisfied. Moreover,
let us assume that for any $t \in \bR$ the values $\gk_j(t)$
are pairwise different. In this case one has
$\bigcap_{t\in\kI}\dom(H(t)) =\{0\}$ for any bounded open interval
$\kI \subseteq  \bR$. Nevertheless, the sets $(I + H_\kI)^{-1}H^{1,2}_a(\kI,\gotH)$
and $(I + H_\kI)^{-1}H^{1,2}_b(\kI,\gotH)$ are dense in $H^{1,2}_a(\kI,\gotH^+_0)$
and $H^{1,2}_b(\kI,\gotH^+_0)$, respectively !

\subsection{Moving potentials}

In this section we consider an example, which is more involved than that we studied above.
\rd{Here we consider the Hamiltonian of two \textit{moving point particles}:}
\be\la{6.20}
H(t) = -\frac{1}{2}\frac{d^2}{dx^2} \dotplus
\gk_1(t)\gd(x - x_1(t)) \dotplus \gk_2(t)\gd(x-x_2(t)) , 
\ee
which domain is described by
\bea\la{6.21}
\lefteqn{
\dom(H(t)) := }\\
& &
\left\{f \in H^{1,2}(\bR):
\ba{l}
f' \in H^{1,2}(\bR \setminus \{x_1(t),x_2(t)\}),\\
\left(f'/2\right)(x_1(t)-0) - \left(f'/2\right)(x_1(t)+0) = \gk_1(t)f(x_1(t)),\\
\left(f'/2\right)(x_2(t)-0) - \left(f'/2\right)(x_2(t)+0) = \gk_2(t)f(x_2(t)),
\ea
\right\}
\nonumber
\eea
in the Hilbert space $L^2(\bR)$. In the following we assume that
$\gk_j(\cdot): \bR \longrightarrow \bR_+$ are \textit{continuous
differentiable} functions. Moreover, we suppose that
\be\la{6.22}
x_1(t) < x_2(t)
\ee
for $t \in \bR$. The sesquilinear
form associated with $H(t)$ is given by
\bead
\lefteqn{
\goth_t[f,g] = }\\
& &
\frac{1}{2} \int_\bR f'(x)\overline{g'(x)} dx +
\gk_1(t)f(x_1(t))\overline{g(x_1(t))} +
\gk_2(t)f(x_2(t))\overline{g(x_2(t))},
\nonumber
\eead
$f,g \in \dom(\goth_t) := H^{1,2}(\bR)$. Notice that the
sesquilinear form $\goth_t$ is non-negative.

To handle this case we start with some formal manipulations.
Using the momentum operator $P$,
\bed
Pf = \frac{1}{i} \ \frac{\partial}{\partial x}f(x), \quad f \in \dom(P) :=
H^{1,2}(\bR),
\eed
we get the representation
\bed
\goth_t[f,g] = \tfrac{1}{2}(Pf,Pg) +
\gk_1(t)f(x_1(t))\overline{g(x_1(t))}
+ \gk_2(t)f(x_2(t))\overline{g(x_2(t))},
\eed
$f,g \in H^{1,2}(\bR)$. The momentum operator generates the
right-shift group $S(\gt) := e^{-i\gt P}$, $\gt \in \bR$, acting as
\bed
(S(\gt)f)(x) = f(x-\gt), \quad f \in L^2(\bR), \quad \gt \in \bR.
\eed
Obviously, one has that
\bed
S(\gt)^{-1}H(t)S(\gt) = \tfrac{1}{2}P^2 +
\gk_1(t) \gd(x - x_1(t) + \gt) + \gk_2(t) \gd(x - x_2(t) + \gt).
\eed
In particular, for $y(t): = \frac{1}{2}(x_1(t) + x_2(t))$ we obtain:
\bead
\lefteqn{
H^S(t) := S(y(t))^{-1}H(t)S(y(t)) =  }\\
& &
e^{iy(t)P}H(t)e^{-iy(t)P} =
\frac{1}{2}P^2 + \gk_1(t)\gd(x + x(t)) + \gk_2(t)\gd(x - x(t)) \,,
\nonumber
\eead
where the relative coordinate obeys
\bed
x(t) := \frac{x_2(t) - x_1(t)}{2} > 0, \quad t \in \bR .
\eed
by \eqref{6.22}.
Further, we define the unitary transformations $W(\gth): L^2(\bR)
\longrightarrow L^2(\bR))$, $\gth > 0$,
\bed
W(\gth)f)(x) := \sqrt{\gth}f(\gth x), \quad f \in L^2(\bR) .
\eed
Let $X$ be  multiplication operator $(X f):=x f(x)$ in $L^2(\bR)$. Then
\bed
L = \frac{1}{2}(XP + PX)
\eed
is a so-called \textit{dilation operator}, which is self-adjoint in $L^2(\bR)$. The
operator $iL$ generates \textit{dilation group} given by
\bed
(e^{isL}f)(x) = e^{s/2}f(e^sx), \quad f \in L^2(\bR), \quad s\in \bR .
\eed
Then we obviously get $W(\gth) = e^{i\ln(\gth)L}$, $\gth > 0$ and
\bead
\lefteqn{
W(\gth)^{-1}H^S(t)W(\gth) =}\\
& &
-\frac{\gth^2}{2}\frac{d^2}{dx^2}
+ \gk_1(t)\gth\gd(x + \gth x(t)) +\gk_2(t)\gth\gd(x - \gth x(t)) \, .
\nonumber
\eead
If we set $\gth = {1}/{x(t)}$, then 
\bead
\lefteqn{
H^{SW}(t) := W(1/x(t))^{-1}H^S(t)W(1/x(t)) =
}\\
& &
e^{i\ln(x(t))L}H^S(t)e^{-i\ln(x(t))L} =
\frac{1}{2x(t)^2}P^2 +
\varkappa_1(t)\gd(x + 1) + \varkappa_2(t)\gd(x - 1) \,,
\nonumber
\eead
where
\bed
\varkappa_1(t) := \frac{\gk_1(t)}{x(t)} \quad \mbox{and} \quad
\varkappa_2(t)  := \frac{\gk_2(t)}{x(t)}.
\eed
\rd{Relation between this Hamiltonian and (\ref{6.20}) has the form:}
\bed
H(t) = e^{-iy(t)P}e^{-i\ln(x(t))L}H^{SW}(t)e^{i\ln(x(t))L}e^{iy(t)P}.
\eed

Now we introduce in the Hilbert space $L^2(\bR,\gotH)$, $\gotH := L^2(\bR)$,  the operator
\bed
\rd{(Df)(t,x) = \left(\frac{1}{i} \ \frac{\partial}{\partial t}f\right)(t,x)}, 
\qquad \dom(D) := H^{1,2}(\bR,\gotH).
\eed
The multiplication operator $S := M(S(y(t)))$, $y(t) =
\frac{1}{2}(x_1(t) + x_2(t))$ \rd{(i.e., $(S f)(t,x):= (S(y(t))f)(t,x)= f(t, x - y(t))$, 
see (\ref{mult-oper}))}, defines a unitary operator on
$L^2(\bR,\gotH)$, and we have that
\bed
D^S := S^{-1}D\; S = D - \dot{y}(t)P \,.
\eed
Similarly, the multiplication operator $W := M(W(1/x(t)))$,
$x(t) = \frac{1}{2}(x_2(t) - x_1(t))$, induces a unitary operator on
$L^2(\bR,\gotH)$. We set
\bed
D^{SW} := W^{-1}D^S W.
\eed
\rd{Since the multiplication operator $W = M(e^{-i\ln(x(t))L})$,} by the commutation relation
$LP - PL = iP$ one gets that
\bed
D^{SW} = D - i\frac{\dot{x}(t)}{x(t)}L - i\frac{\dot{y}(t)}{x(t)}P.
\eed
Now we set
\bed
H^{SW} := W^{-1}S^{-1}H\,S\,W
\eed
and
\bed
\widetilde{K}^{SW} := D^{SW} + H^{SW}
\eed
with domain $\dom(\widetilde{K}^{SW}): = \dom(D^{SW}) \cap
\dom(H^{SW})$. Then a straightforward computation gives that this operator is equal to 
\bed
\widetilde{K}^{SW} := D + L_0
\eed
with domain $\dom(\widetilde{K}^{SW}) = \dom(D) \cap \dom(L_0)$, where
\bead
L_0(t) & := &\frac{1}{2x(t)^2}(P - x(t)(\dot{x}(t)X + \dot{y}(t))^2 -
\frac{1}{2}(\dot{x}(t)X + \dot{y}(t))^2\\
           &    &  + \varkappa_1(t)\gd(x + 1) + \varkappa_2(t)\gd(x - 1) \,.\nonumber
\eead
Finally, let us introduce the gauge transformation
\bed
(\gG(t)f)(x) := e^{{i}\int^t_0\left((\dot{x}(s)x + \dot{y}(s))^2
    + x^2\right)ds/2}f(x), \quad f \in L^2(\bR),
\eed
which induces the multiplication operator $\gG :=M(\gG(t))$ on $L^2(\kI,\gotH)$. Then we find
\bed
\widetilde{\bf K} :=
\widetilde{K}^{SW\gG} := \gG^{-1}K^{SW}\gG = D + L \,,
\eed
where operator
\bead
\lefteqn{
L(t) := }\\
& &
\frac{1}{2x(t)^2}(P + \gb_1(t)X + \gb_0(t))^2 + \frac{1}{2}X^2 +
\varkappa_1(t)\gd(x + 1) + \varkappa_2(t)\gd(x - 1)
\nonumber
\eead
with
\bed
\gb_1(t) := \int^t_0 (\dot{x}(s)^2 + 1)ds - x(t)\dot{x}(t)
\eed
and
\bed
\gb_0(t) := \int^t_0 \dot{y}(s)\dot{x}(s)ds - x(t)\dot{y}(t).
\eed
\rd{As above the family $\{L(t)\}_{t\in\bR}$,
is measurable and defines a densely defined self-adjoint
multiplication operator $L:=M(L(t))$ on $L^2(\kI,\gotH)$.}
Then the operators $\widetilde {K} := D + H$ and
$\widetilde{\bf K} = D + L$ are related by
\be\la{6.50}
\widetilde{K} = S\,W\,\gG\,\widetilde{\bf K}\,\gG^{-1}\,W^{-1}\,S^{-1} \,.
\ee

Instead to solve  the bidirectional evolution equation \eqref{5.28}
we consider the modified bidirectional evolution equation
\be\la{6.51}
\frac{1}{i} \ \frac{\partial}{\partial t}u(t) + L(t)u(t) = 0.
\ee
Following Section 5 we introduce the family of quadratic forms $\gotl_t[\cdot,\cdot]$
\bead
\gotl_t[f,g] & := &
\frac{1}{2x(t)^2}(Pf + \gb_1(t)Xf + \gb_0(t)f,Pf +
\gb_1(t)Xg + \gb_0(t)g) + \nonumber\\
& &
\frac{1}{2}(Xf,Xg) + \varkappa_1(t)f(-1)\overline{g(-1)} +
\varkappa_2(t)f(1)\overline{g(1)} + (f,g),
\nonumber
\eead
$f,g \in \dom(\gotl_t) := \dom(P) \cap \dom(X)$ corresponding to operators $L(t)$,
and define the norm
\bed
\|f \|^+_t := \|\sqrt{L(t) + I}f\| = \sqrt{\gotl_t[f,f] + \|f\|^2},
\eed
$f \in \dom(\sqrt{L(t) + I}) = \dom(\gotl_t) =
\dom(P) \cap \dom(X)$. It is easy to check that the domain
$\dom(\gotl_t)$ is independent of $t \in \bR$. By $\gotL^+_t$ we
denote the Hilbert space, which arises when we endow the domain
$\dom(\gotl_t)$ with the scalar product $(f,g)^+_t := \gotl_t[f,g] +(f,g)$.
Note that the norm $\|\cdot\|^+_t$ is
equivalent to the norm $\|f\|_{PX} = \sqrt{\|Pf\|^2 + \|Xf\|^2}$, $f \in \dom(P) \cap \dom(X)$.

\rd{Now we proceed as in the previous section.} First we find
\bead
\lefteqn{
\frac{d}{dt}(\|f\|^+_t)^2 =
\frac{\dot{x}(t)}{x(t)^3}\|Pf + \gb_1(t)Xf + \gb_0(t)f\|^2}\\
& &
+ \frac{2}{x(t)^2}\real(Pf + \gb_1(t)Xf + \gb_0(t)f,\dot{\gb}_1(t)Xf + \dot{\gb}_0(t)f) \nonumber\\
& &
+ \dot{\varkappa}_1(t)|f(-1)|^2 + \dot{\varkappa}_2(t)|f(1)|^2.
\nonumber
\eead
A straightforward computation shows that for any bounded
interval $\kI$ there is a constant $\gga_\kI$
such that
\bed
\left|\frac{d}{dt}(\|f\|^+_t)^2\right| \le 2\gga_\kI(\|f\|^+_t)^2
\eed
for $t \in \kI$ which yields
\bed
-\gga_\kI \le \frac{d}{dt} \ln(\|f\|^+_t) \le \gga_\kI.
\eed
Hence we obtain the estimate:
\bed
-\gga_\kI(t-s) \le \ln(\|f\|^+_t) - \ln(\|f\|^+_s) \le
\gga_\kI(t-s)
\eed
for $t,s \in \kI$ and $s \le t$,  which yields
\bed
\|f\|^+_t \le e^{\gga_\kI(t-s)}\|f\|^+_s, \quad t,s \in \kI,
\quad s\le t.
\eed
The last relation implies \eqref{5.23}. By virtue of Theorem \ref{V.2} we
get that the \textit{auxiliary}  bidirectional evolution equation
\bed
\frac{\partial}{\partial t}u(t) + iL^-(t)u(t) = 0
\eed
admits a unique solution $\{\Lambda^-(t,s)\}_{(t,s) \in \bR \times \bR}$ on $\bR$.
By Theorem \ref{V.4} the \textit{original}
bidirectional evolution equation \eqref{6.51} admits a solution for
which the Hilbert space $\gotL^+_0$ is admissible.
By the same line of reasoning as for non-moving point interactions one can prove
that there is unique unitary solution $\{\gL(t,s)\}_{(t,s) \in \bR\times\bR}$
of the bidirectional evolution
equation \eqref{6.51} for which $\gotL^+_0$ is admissible.

These results allow to prove that the original forward evolution equation \eqref{5.21} on $\bR$
admits a solution. To this end one has to verify that for any bounded interval $\kI$ the extension 
of the forward generator ${\bf K}_\kI$ of $\widetilde{\bf K}_\kI$ defines
an extension of the forward generator $K_\kI$ of $\widetilde{K}_\kI$
defined by $K_\kI := S\;W\;\gG \;{\bf K}_\kI\;\gG^{-1}\;W^{-1}\;S^{-1}$.
\rd{However, this is evident since it follows form the representation \eqref{6.50}.}
Similarly, one proves that for any bounded interval $\kI$ the backward
generator extension ${\bf K}^\kI$ of $\widetilde{\bf K}^ \kI$
defines a backward generator extension $K^\kI$ of $\widetilde{K}^\kI$
by $K^\kI := S\;W\;\gG \;{\bf K}^\kI\;\gG^{-1}\;W^{-1}\;S^{-1}$.
By these we immediately obtain that the bidirectional propagator 
$\{G(t,s)\}_{(t,s) \in \bR \times\bR}$ defined by
\be\la{6.60}
G(t,s) :=
e^{-iy(t)P}e^{-i\ln(x(t))L}\gG(t)\gL(t,s)\gG(s)^{-1}e^{i\ln(x(s))L}e^{iy(s)P},
\ee
for any $(t,s) \in \bR \times \bR$, is a solution of the bidirectional evolution equation
\eqref{5.28}.

It remains only to identify the subspace which is admissible with respect
to $\{G(t,s)\}_{(t,s) \in \bR\times\bR}$.
We recall that $\gotL^+_0$ is the subspace which is admissible with
respect to $\{\Lambda(t,s)\}_{(t,s)\in \bR \times \bR}$.
If we set
\bed
H^\gG(t) := \gG(t)L(t)\gG(t)^{-1}, \quad t \in \bR \,,
\eed
then a straightforward computation shows that
\bead
H^\gG(t) & := & \frac{1}{2x(t)^2}(P -x(t)\dot{x}(t)X - x(t)\dot{y}(t))^2 + \\
& & \frac{1}{2}X^2 + \varkappa_1(t)(t)\gd(x + 1) + \varkappa_2(t)\gd(x-1).
\eead
Further, setting
\bed
H^{\gG W}(t) := e^{-i\ln(x(t))L}\gG(t) L(t)\gG(t)^{-1}e^{i\ln(x(t))L}
\eed
we find that 
\bed
H^{\gG W}(t) =
\frac{1}{2}(P - \frac{\dot{x}(t)}{x(t)}X - \dot{y}(t))^2 +
\frac{1}{2x(t)^2}X^2 +
\gk_1(t)\gd(x + x(t)) + \gk_2(t)\gd(x-x(t)).
\eed
Finally, we introduce the family:
\bed
H^{\gG WS}(t) :=
e^{-iy(t)P}e^{-i\ln(x(t))L}\gG(t)L(t))\gG(t)^{-1}e^{i\ln(x(t))L}e^{iy(t)P}
\eed
\rd{which implies} 
\bead
\lefteqn{
H^{\gG WS}(t) = \frac{1}{2}\left(P - \frac{\dot{x}(t)}{x(t)}(X
  - y(t))\right)^2 +}\\
& &
\frac{1}{2x(t)^2}(X - y(t))^2 +
\gk_1(t)\gd(x - x_1(t)) + \gk_2(t)\gd(x-x_2(t)).
\eead
\rd{For a shorthand  let $Z(t):=H^{\gG WS}(t)$.} Then quadratic form associated with $Z(t)$
we denote by $\gotz_t[\cdot,\cdot]$. One can easily verify that
the domain $\dom(\gotz_t)$ is independent of $t \in \bR$. The Hilbert space which is associated with
$\gotz_t$ is denoted by $\gotZ^+_t$. A straightforward computation
shows that for any $t \in \bR$ the Hilbert space $\gotZ^+_t$ can be identified with
$\gotH_{PX} := \{\dom(P) \cap \dom(X),\|\cdot\|_{PX}\}$.
It is obvious, that the Hilbert space $\gotH_{PX}$ is admissible for the
bidirectional propagator $\{G(t,s)\}_{(t,s)\in \bR}$ defined by
\eqref{6.60}.
Summing up one gets the following theorem:
\bt
Let $\gk_j(\cdot): \bR \longrightarrow \bR_+$ and $x_j(\cdot): \bR
\longrightarrow \bR$ be continuously differentiable
functions. Further, let $\{H(t)\}_{t\in\bR}$ be the family of
non-negative self-adjoint operators given by \eqref{6.20} and
\eqref{6.21}. If the condition \eqref{6.22} is satisfied for any $t
\in \bR$, then the bidirectional evolution equation
\eqref{5.28} is well-posed on $\bR$ for $p=2$ and
possesses a unique unitary solution for which the Hilbert space
$\gotH_{PX}$ is admissible.
\et

\bigskip

\noindent{\rd{\textbf{Acknowledgments}}}

\rd{This papers results from numerous discussions and mutual visits of the co-authors 
to Berlin and Marseille.} \gn{H.N. thanks the CPT (Marseille) for
hospitality and financial support.} 
\rd{V.A.Z. is thankful to WIAS (Berlin) for invitations and financial support of his several
visits.}



\begin{thebibliography}{10}

\bibitem{HiPhi}
{\em Functional analysis and semi-groups}.
\newblock American Mathematical Society Colloquium Publications, vol. 31.
  American Mathematical Society, Providence, R. I., 1957.
\newblock rev. ed.

\bibitem{CM}
Pilar Cembranos and Jos{\'e} Mendoza.
\newblock {\em Banach spaces of vector-valued functions}, volume 1676 of {\em
  Lecture Notes in Mathematics}.
\newblock Springer-Verlag, Berlin, 1997.

\bibitem{Con1}
Adrian Constantin.
\newblock The construction of an evolution system in the hyperbolic case and
  applications.
\newblock {\em Math. Nachr.}, 224:49--73, 2001.

\bibitem{Gr1}
G.~Da~Prato and P.~Grisvard.
\newblock Sommes d'op\'erateurs lin\'eaires et \'equations diff\'erentielles
  op\'erationnelles.
\newblock {\em J. Math. Pures Appl. (9)}, 54(3):305--387, 1975.

\bibitem{DFT2}
G.~F. Dell'Antonio, R.~Figari, and A.~Teta.
\newblock A limit evolution problem for time-dependent point interactions.
\newblock {\em J. Funct. Anal.}, 142(1):249--275, 1996.

\bibitem{DFT1}
G.~F. Dell'Antonio, R.~Figari, and A.~Teta.
\newblock The {S}chr\"odinger equation with moving point interactions in three
  dimensions.
\newblock In {\em Stochastic processes, physics and geometry: new interplays, I
  (Leipzig, 1999)}, volume~28 of {\em CMS Conf. Proc.}, pages 99--113. Amer.
  Math. Soc., Providence, RI, 2000.

\bibitem{D1}
Gianfausto Dell'Antonio.
\newblock Point interactions.
\newblock In {\em Mathematical physics in mathematics and physics (Siena,
  2000)}, volume~30 of {\em Fields Inst. Commun.}, pages 139--150. Amer. Math.
  Soc., Providence, RI, 2001.

\bibitem{DU1}
J.~Diestel and J.~J. Uhl, Jr.
\newblock {\em Vector measures}.
\newblock American Mathematical Society, Providence, R.I., 1977.
\newblock Mathematical Surveys, No. 15.

\bibitem{DoFr1}
John~D. Dollard and Charles~N. Friedman.
\newblock Asymptotic behavior of solutions of linear ordinary differential
  equations.
\newblock {\em J. Math. Anal. Appl.}, 66(2):394--398, 1978.

\bibitem{Do1}
J.~R. Dorroh.
\newblock A simplified proof of a theorem of {K}ato on linear evolution
  equations.
\newblock {\em J. Math. Soc. Japan}, 27(3):474--478, 1975.

\bibitem{El1}
Joanne Elliott.
\newblock The equation of evolution in a {B}anach space.
\newblock {\em Trans. Amer. Math. Soc.}, 103:470--483, 1962.

\bibitem{Ev1}
David~E. Evans.
\newblock Time dependent perturbations and scattering of strongly continuous
  groups on {B}anach spaces.
\newblock {\em Math. Ann.}, 221(3):275--290, 1976.

\bibitem{Go1}
Jerome~A. Goldstein.
\newblock Time dependent hyperbolic equations.
\newblock {\em J. Functional Analysis}, 4:31--49, 1969.

\bibitem{Go2}
Jerome~A. Goldstein.
\newblock On the absence of necessary conditions for linear evolution
  operators.
\newblock {\em Proc. Amer. Math. Soc.}, 64(1):77--80, 1977.

\bibitem{Ha1}
Matthew Hackman.
\newblock The abstract time-dependent {C}auchy problem.
\newblock {\em Trans. Amer. Math. Soc.}, 133:1--50, 1968.

\bibitem{Hey1}
Eugen Heyn.
\newblock Die {D}ifferentialgleichung {$dT/dt=P(t)T$} {$f\ddot ur$}
  {O}peratorfunktionen.
\newblock {\em Math. Nachr.}, 24:281--330, 1962.

\bibitem{How}
James~S. Howland.
\newblock Stationary scattering theory for time-dependent {H}amiltonians.
\newblock {\em Math. Ann.}, 207:315--335, 1974.

\bibitem{Is1}
Susumu Ishii.
\newblock An approach to linear hyperbolic evolution equations by the {Y}osida
  approximation method.
\newblock {\em Proc. Japan Acad. Ser. A Math. Sci.}, 54(1):17--20, 1978.

\bibitem{Ka2}
Tosio Kato.
\newblock Integration of the equation of evolution in a {B}anach space.
\newblock {\em J. Math. Soc. Japan}, 5:208--234, 1953.

\bibitem{Ka3}
Tosio Kato.
\newblock On linear differential equations in {B}anach spaces.
\newblock {\em Comm. Pure Appl. Math.}, 9:479--486, 1956.

\bibitem{Ka1}
Tosio Kato.
\newblock {\em Perturbation theory for linear operators}.
\newblock Die Grundlehren der mathematischen Wissenschaften, Band 132.
  Springer-Verlag New York, Inc., New York, 1966.

\bibitem{Ka6}
Tosio Kato.
\newblock Linear evolution equations of ``hyperbolic'' type.
\newblock {\em J. Fac. Sci. Univ. Tokyo Sect. I}, 17:241--258, 1970.

\bibitem{Ka7}
Tosio Kato.
\newblock Linear evolution equations of ``hyperbolic'' type. {II}.
\newblock {\em J. Math. Soc. Japan}, 25:648--666, 1973.

\bibitem{Ki1}
J.~Kisy{\'n}ski.
\newblock Sur les op\'erateurs de {G}reen des probl\`emes de {C}auchy
  abstraits.
\newblock {\em Studia Math.}, 23:285--328, 1963/1964.

\bibitem{Ko1}
Kazuo Kobayasi.
\newblock On a theorem for linear evolution equations of hyperbolic type.
\newblock {\em J. Math. Soc. Japan}, 31(4):647--654, 1979.

\bibitem{Kom1}
Yukio K{\=o}mura.
\newblock On linear evolution operators in reflexive {B}anach spaces.
\newblock {\em J. Fac. Sci. Univ. Tokyo Sect. I A Math.}, 17:529--542, 1970.

\bibitem{Kr1}
S.~G. Kre{\u\i}n.
\newblock {\em Lineinye uravneniya v banakhovom prostranstve}.
\newblock Vorone\v z. Gosudarstv. Univ., Voronezh, 1968.

\bibitem{LM1}
Yuri Latushkin and Stephen Montgomery-Smith.
\newblock Evolutionary semigroups and {L}yapunov theorems in {B}anach spaces.
\newblock {\em J. Funct. Anal.}, 127(1):173--197, 1995.

\bibitem{Mi1}
Sigeru Mizohata.
\newblock Le probl\`eme de {C}auchy pour les \'equations paraboliques.
\newblock {\em J. Math. Soc. Japan}, 8:269--299, 1956.

\bibitem{MRh1}
Sylvie Monniaux and Abdelaziz Rhandi.
\newblock Semigroup methods to solve non-autonomous evolution equations.
\newblock {\em Semigroup Forum}, 60(1):122--134, 2000.

\bibitem{N0}
H.~Neidhardt.
\newblock {\em {Integration of Evolutionsgleichungen mit Hilfe von
  Evolutionshalbgruppen}}.
\newblock Dissertation, AdW der DDR.
\newblock Berlin 1979.

\bibitem{N1}
Hagen Neidhardt.
\newblock On abstract linear evolution equations. {I}.
\newblock {\em Math. Nachr.}, 103:283--298, 1981.

\bibitem{N2}
Hagen Neidhardt.
\newblock {On abstract linear evolution equations. II.}
\newblock Technical report, {Prepr., Akad. Wiss. DDR, Inst. Math.
  P-MATH-07/81}, Berlin, 1981.

\bibitem{N3}
Hagen Neidhardt.
\newblock {On linear evolution equations. III: Hyperbolic case.}
\newblock Technical report, {Prepr., Akad. Wiss. DDR, Inst. Math.
  p-MATH-05/82}, Berlin, 1982.

\bibitem{Ni1}
Gregor Nickel.
\newblock Evolution semigroups and product formulas for nonautonomous {C}auchy
  problems.
\newblock {\em Math. Nachr.}, 212:101--116, 2000.

\bibitem{NiSch1}
Gregor Nickel and Roland Schnaubelt.
\newblock An extension of {K}ato's stability condition for nonautonomous
  {C}auchy problems.
\newblock {\em Taiwanese J. Math.}, 2(4):483--496, 1998.

\bibitem{Ph1}
R.~S. Phillips.
\newblock Perturbation theory for semi-groups of linear operators.
\newblock {\em Trans. Amer. Math. Soc.}, 74:199--221, 1953.

\bibitem{Po1}
Andrea Posilicano.
\newblock The {S}chr\"odinger equation with a moving point interaction in three
  dimensions.
\newblock {\em Proc. Amer. Math. Soc.}, 135(6):1785--1793 (electronic), 2007.

\bibitem{RRhSch1}
Frank R{\"a}biger, Abdelaziz Rhandi, and Roland Schnaubelt.
\newblock Perturbation and an abstract characterization of evolution
  semigroups.
\newblock {\em J. Math. Anal. Appl.}, 198(2):516--533, 1996.

\bibitem{RSch1}
Frank R{\"a}biger and Roland Schnaubelt.
\newblock The spectral mapping theorem for evolution semigroups on spaces of
  vector-valued functions.
\newblock {\em Semigroup Forum}, 52(2):225--239, 1996.

\bibitem{SY2}
M.~R. Sayapova and D.~R. Yafaev.
\newblock Scattering theory for potentials of zero radius which are periodic
  with respect to time.
\newblock In {\em Spectral theory. Wave processes}, volume~10 of {\em Probl.
  Mat. Fiz.}, pages 252--266, 301. Leningrad. Univ., Leningrad, 1982.

\bibitem{SY1}
M.~R. Sayapova and D.~R. Yafaev.
\newblock The evolution operator for time-dependent potentials of zero radius.
\newblock {\em Trudy Mat. Inst. Steklov.}, 159:167--174, 1983.
\newblock Boundary value problems of mathematical physics, 12.

\bibitem{Ta1}
Hiroki Tanabe.
\newblock {\em Equations of evolution}, volume~6 of {\em Monographs and Studies
  in Mathematics}.
\newblock Pitman (Advanced Publishing Program), Boston, Mass., 1979.

\bibitem{Tan1}
Naoki Tanaka.
\newblock Generation of linear evolution operators.
\newblock {\em Proc. Amer. Math. Soc.}, 128(7):2007--2015, 2000.

\bibitem{Tan2}
Naoki Tanaka.
\newblock A characterization of evolution operators.
\newblock {\em Studia Math.}, 146(3):285--299, 2001.

\bibitem{Tan3}
Naoki Tanaka.
\newblock Nonautonomous abstract {C}auchy problems for strongly measurable
  families.
\newblock {\em Math. Nachr.}, 274/275:130--153, 2004.

\bibitem{Tr1}
H.~F. Trotter.
\newblock On the product of semi-groups of operators.
\newblock {\em Proc. Amer. Math. Soc.}, 10:545--551, 1959.

\bibitem{Y1}
D.~R. Yafaev.
\newblock Scattering theory for time-dependent zero-range potentials.
\newblock {\em Ann. Inst. H. Poincar\'e Phys. Th\'eor.}, 40(4):343--359, 1984.

\bibitem{Ya1}
Atsushi Yagi.
\newblock On a class of linear evolution equations of ``hyperbolic'' type in
  reflexive {B}anach spaces.
\newblock {\em Osaka J. Math.}, 16(2):301--315, 1979.

\bibitem{Ya2}
Atsushi Yagi.
\newblock Remarks on proof of a theorem of {K}ato and {K}obayasi on linear
  evolution equations.
\newblock {\em Osaka J. Math.}, 17(1):233--243, 1980.

\bibitem{Yo1}
K{\^o}saku Yosida.
\newblock Time dependent evolution equations in a locally convex space.
\newblock {\em Math. Ann.}, 162:83--86, 1965/1966.

\bibitem{Yo2}
K{\^o}saku Yosida.
\newblock {\em Functional analysis}.
\newblock Second edition. Die Grundlehren der mathematischen Wissenschaften,
  Band 123. Springer-Verlag New York Inc., New York, 1968.

\end{thebibliography}
\end{document}